\documentclass[]{elsarticle}

\usepackage{amsmath}

\biboptions{authoryear}
\setcitestyle{square}
\setcitestyle{comma}
\setcitestyle{numbers}

\usepackage{color}

\usepackage{array}
\usepackage{csquotes}
\usepackage{enumitem}
\usepackage{amssymb}
\usepackage[colorlinks,urlcolor=blue]{hyperref}
{\left\lbrace\begin{array}{@{}l@{}}}%
{\end{array}\right.}

\begin{document}

\begin{frontmatter}

\title{Probabilistic buckling of imperfect hemispherical shells containing a distribution of defects}


\author[mymainaddress]{Fani Derveni\fnref{T}}
\ead{fani.derveni@epfl.ch}

\author[mymainaddress]{William Gueissaz\fnref{T}}
\ead{william.gueissaz@epfl.ch}
\fntext[T]{FD and WG have contributed equally to this work.}

\author[mymainaddress]{Dong Yan}
\ead{dong.yan@epfl.ch}

\author[mymainaddress]{Pedro M. Reis\corref{mycorrespondingauthor}}
\cortext[mycorrespondingauthor]{Corresponding author}
\ead{pedro.reis@epfl.ch}

\address[mymainaddress]{\'{E}cole Polytechnique F\'{e}d\'{e}rale de Lausanne (EPFL), Flexible Structures Laboratory, CH-1015 Lausanne, Switzerland\\
    }

\begin{abstract}
The buckling of spherical shells is plagued by a strong sensitivity to imperfections. Traditionally, imperfect shells tend to be characterized empirically by the knockdown factor, the ratio between the measured buckling strength and the corresponding classic prediction for a perfect shell. Recently, it has been demonstrated that the knockdown factor of a shell containing a single imperfection can be predicted when there is detailed \textit{a priori} knowledge of the defect geometry. Still, addressing the analogous problem for a shell containing many defects remains an open question. Here, we use finite element simulations, which we validate against precision experiments, to investigate hemispherical shells containing a well-defined distribution of imperfections. Our goal is to characterize the resulting knockdown factor statistics. First, we study the buckling of shells containing only two defects, uncovering nontrivial regimes of interactions that echo existing findings for cylindrical shells. Then, we construct statistical ensembles of imperfect shells, whose defect amplitudes are sampled from a lognormal distribution. We find that a 3-parameter Weibull distribution is an excellent description for the measured statistics of knockdown factors, suggesting that shell buckling can be regarded as an extreme-value statistics phenomenon.
\end{abstract}

\begin{keyword}
Probabilistic buckling \sep Random imperfections \sep Knockdown factors \sep Experiments \sep Finite element modeling \sep Micro-computed tomography.
\end{keyword}

\end{frontmatter}

\newpage
\section{Introduction}
\label{sec:introduction}
Thin shell structures with their inherent curved, slender configurations, have been studied extensively across a wide range of length scales, from viruses~\cite{lidmar2003virus}, pollen grains~\cite{katifori_foldable_2010}, and plants~\cite{yin_stress-driven_2008}, to drug-delivery capsules~\cite{datta_delayed_2012}, gas tanks~\cite{godoy_buckling_2016}, and aeronautical structures~\cite{hoff1966thin}. Due to their slenderness, shells are prone to buckling, whose subcritical nature can cause the catastrophic loss of load-bearing capacity and sudden changes in the deformation mode~\cite{pearson_collapse_2006}. Whereas shell buckling is often regarded as a phenomenon to avoid in traditional structural mechanics, recent studies have enlightened the benefits of leveraging the buckled configurations in attaining advantageous mechanical properties~\cite{djellouli_buckling_2017,gorissen_inflatable_2020,sacanna_lock_2011}. These studies reflect the change in perspective from \textit{buckliphobia} to \textit{buckliphilia}~\cite{reis2015perspective}.

For the design of shell structures, it is desirable to have accurate predictions for their limits of stability. The pioneering work of Zoelly~\cite{zoelly} provides the widely used prediction for the critical buckling condition of a pressurized perfect spherical shell:
\begin{align}\label{eqn:zolley}
p_\mathrm{c} = \frac{2E}{\sqrt{3(1-\nu^2)}}\eta^{-2},
\end{align}
where $E$ is the Young's modulus, $\nu$ is the Poisson's ratio, and $\eta = R/t$ is the radius ($R$) to thickness ($t$) ratio, measuring the slenderness of the spherical shell. Subsequently, a large body of experimental studies has shown that Eq.~(\ref{eqn:zolley}) systematically overpredicts experimental measurements~\cite{tsien_theory_1942,kaplan_nonlinear_1954,homewood1961experimental,seaman1962nature, kiernan_elastic_1963,carlson1967experimental}, due to a strong sensitivity of the buckling to material and/or geometric imperfections. Koiter~\cite{koiter_over_1945} proposed a general theory of elastic stability to study the postbuckling behavior of structures. Instigated by this seminal work, much theoretical effort has been dedicated to study shell buckling in the presence of assumed imperfections in geometry, loading, and boundary conditions~\cite{hutchinson_effect_1971,budiansky_buckling_1972,Bijlaard_elastic_1960, kobayashi_influence_1968, hutchinson2016buckling}.
Still, given the lack of theoretical bases to predict their critical buckling strength, realistic (imperfect) shells tend to be characterized empirically by the \textit{knockdown factor}, 
\begin{align}\label{eqn:knockdown_def}
\kappa = \frac{p_\mathrm{max}}{p_\mathrm{c}},
\end{align}
where $p_\mathrm{max}$ is the measured maximum buckling pressure supported by the shell and $p_\mathrm{c}$ is the corresponding classic prediction from Eq.~(\ref{eqn:zolley}) if the shell were to be perfect. For example, NASA has largely relied on knockdown factors in the characterization of shell structures~\cite{seide1960development, peterson1968buckling}, using vast experimental datasets to devise empirical guidelines to aid the design of imperfection-sensitive shells.

Significant progress in the experimental study of shell buckling was made recently, in part due to the introduction of a versatile shell fabrication technique proposed by Lee \emph{et al.}~\cite{lee_fabrication_2016}. By coating a rigid hemispherical mold with a liquid polymer solution, this technique yields a shell of nearly uniform thickness upon curing. Lee \emph{et al.}~\cite{lee_geometric_2016} then used a flexible mold, deformed by an indenter applied at the pole, to produce shells containing a single, precisely engineered, dimple-like defect. Importantly, this technique by Lee \emph{et al.}~\cite{lee_geometric_2016} offers precise and systematic control over the geometry of the engineered defect, enabling accurate and thorough measurements of the knockdown factor as a function of the imperfections characteristics.
Their experimental results evidenced that, as the amplitude of a single dimple-like defect increases, the knockdown factor decreases rapidly until it reaches a plateau, the onset of which occurs when the defect amplitude is approximately equal to the shell thickness. The authors also demonstrated that when the geometry of the imperfection is known in detail, the knockdown factor can be predicted accurately, either through finite element modeling or reduced-order shell theory models. The dependence of the knockdown factor and its plateau on the geometric parameters of the defect and the shell has also been studied in detail~\cite{jimenez2017technical}. All of these results for imperfect spherical shells are qualitatively consistent with much earlier theoretical analyses for cylindrical shells~\cite{budiansky_buckling_1972}, but this time with accurate quantitative predictions validated by precision experiments. The case of a through-thickness single imperfection on a hemispherical shell has also been studied~\cite{yan_buckling_2020}.

All the aforementioned studies measured the knockdown factor by buckling the imperfect shells. As an alternative, a non-destructive poking technique was proposed in Refs.~\cite{thompson_probing_2017,virot_stability_2017} to probe the critical buckling load of an imperfect cylindrical shell from its response to indentation at different axial loading levels to extrapolate the critical conditions without actually breaking the shell. This technique has also been successfully applied to spherical shells~\cite{marthelot2017buckling,abbasi_probing_2021}. Further experimental and numerical studies revealed that, due to the localized deformation caused by the defect, the reliability of the poking test is strongly affected by the distance between the poking location and the imperfections, for both cylindrical~\cite{fan2019critical,yadav2021nondestructive} and spherical shells~\cite{marthelot2017buckling,abbasi_probing_2021}. Particularly, Fan \emph{et al.}~\cite{fan2019critical} performed numerical poking tests on shells with two dimple-like defects and with a distribution of defects measured from an aluminum can. They showed that the poker had to be positioned sufficiently close to the center of the most deleterious defect in order to yield accurate predictions.

Despite the significant advances mentioned above in theoretical, experimental, and computational studies of shells containing a single imperfection, there remain many open questions in the more realistic and practically relevant case of shell buckling due to multiple, or even a large distribution of imperfections, potentially influenced by defect-defect interactions. Wullschleger~\cite{wullschleger2006numerical} studied imperfect cylindrical shells containing two defects at varying levels of separation, and showed that the interactions become important when the defects are close to each other. Beyond the two-defect case, various probabilistic methods to predict the knockdown factor of cylindrical shells have been proposed. Axisymmetric defects on cylindrical shells were investigated in the earlier work of Amazigo~\cite{amazigo1969buckling} via a modified truncated hierarchy method, concluding that the buckling capacity  depends on the spectral imperfection density. Cylindrical shells with random imperfections subjected to axial compression have also been examined using the Monte Carlo method, imposing either axisymmetric~\cite{elishakoff1982reliability} or asymmetric~\cite{elishakoff1985reliability} defects. These studies established probabilistic methods as the more suitable approach to assess the design criteria for cylindrical shells when compared with deterministic approaches. These various probabilistic methods relevant for shell buckling have been reviewed by Elishakoff~\cite{elishakoff_probabilistic_2012}. However, for spherical shells, to the best of our knowledge, the influence of defect-defect interactions on their buckling behavior has not been reported to date. Furthermore, statistics of the knockdown factor for spherical shells containing distributions of defects remain mostly unexplored. 

Here, we investigate the buckling of imperfect hemispherical shells containing a large number of defects, and compare the statistics of knockdown factors to the classical case of single-defect shell buckling. Our study follows two stages, combining experiments and simulations.
First, we fabricate elastomeric hemispherical shells using 3D-printed molds containing several imperfections and a polymer-coating technique. We characterize the full 3D geometry of the experimental specimens using X-ray micro-computed tomography ($\mu$CT). Finally, we obtain knockdown factors via physical buckling tests, which are then used to validate the Finite Element Method (FEM) simulations.
Second, we turn our attention to the validated FEM and study the two-defect case where the distance between the two defects is systematically varied, uncovering a possible interaction regime. Then, we construct statistical ensembles of shells, each containing a large number of defects whose amplitude is treated as a lognormal distributed random variable. By simulating the buckling of these ensembles of imperfect shells using FEM, we compute the knockdown factor statistics. Our results show that the probability density function (PDF) of the knockdown factor is described well by a Weibull distribution. We also find that the mode (peak) of the knockdown factor PDF, decreases when both the mean defect amplitude and its standard deviation increase. In addition, when the minimum distance between defects is large, we observe that the knockdown factors are exclusively dominated by the largest defect, similar to what is observed in the single-defect case. However, when the minimum separation decreases, defect-defect interactions play a significant role in dictating the knockdown factor. Our findings suggest that shell buckling can be placed in the broader class of extreme-value statistics phenomena, calling for the study of probabilistic shell buckling in even more practically relevant imperfection conditions.

Our paper is structured as follows. In Section~\ref{sec:problem_definition}, we define the problem and state the main research questions. Section~\ref{sec:methods_experiments} presents the experimental methodology, including (a) the protocol to fabricate the imperfect shells, (b) the characterization of their geometry using X-Ray micro-computed tomography, and (c) the experimental pressure-buckling tests. Section~\ref{sec:methods_FEM} details the FEM simulations, which are then validated against experiments in Section~\ref{sec:validation}. Section~\ref{sec:results_2defects} is dedicated to the case of an imperfect shell containing two defects. Results of the buckling statistics for shells containing a distribution of imperfections are presented in Section~\ref{sec:results_distribution}.
Section~\ref{sec:conclusion} provides the conclusions of our work, including a discussion of our findings and recommendations for future work.
\section{Definition of the problem}
\label{sec:problem_definition}

\begin{figure}[h]
    \centering\includegraphics[width=0.95\textwidth]{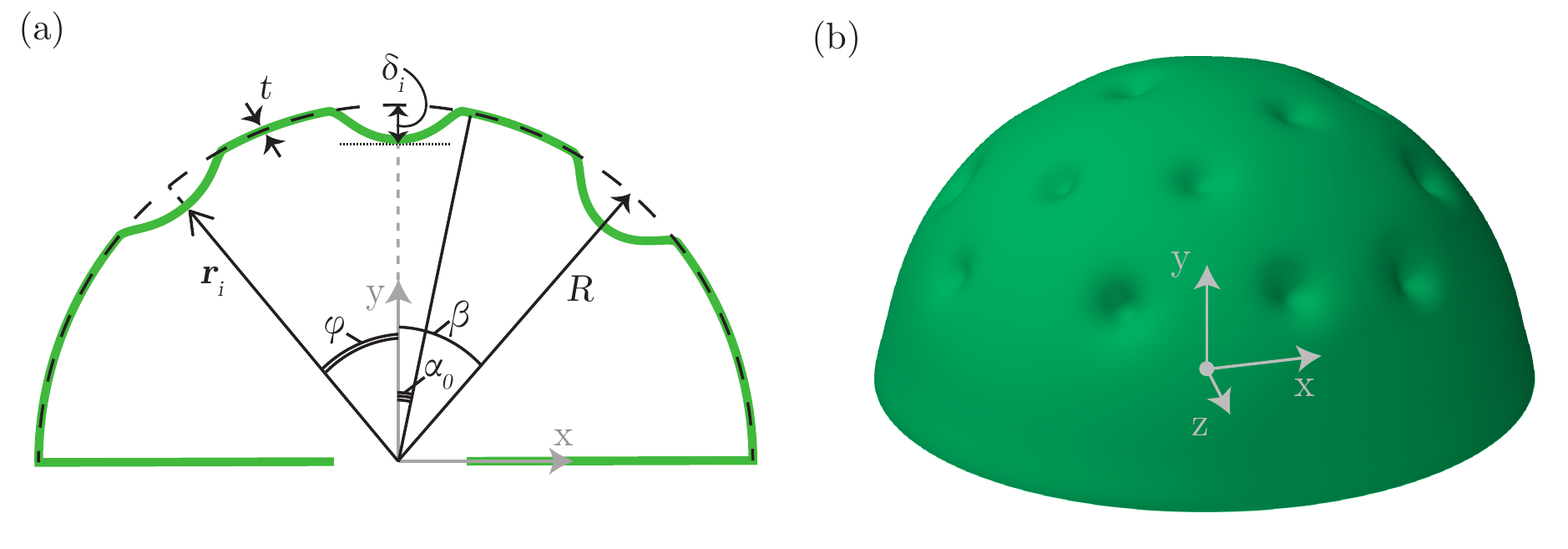}
    \caption{Problem Definition. (a) A two-dimensional schematic of an elastic hemispherical shell with three dimple defects.
    (b) A three-dimensional representation of a shell with multiple dimple imperfections, $w_{i}$ (Eq.~\ref{eqn:geom_gaussiandimple}), whose amplitudes, $\overline\delta_i$, follow a lognormal distribution (Eq.~\ref{eqn:lognormal}).
    }
    \label{fig:probdef}
    \end{figure}

We consider a thin, elastic, and hemispherical shell containing a distribution of geometric imperfections, as illustrated in
Fig.~\ref{fig:probdef}. The overall geometry of the shell is described by its nominal radius, $R$, and thickness, $t$, with a radius-to-thickness ratio $\eta=R/t$. The shell contains $N>1$ (typically $N\gg1$) defects, each of which is assigned an index $i$. The  $i$-th defect introduces a radial deviation, $w_i$, of the otherwise spherical mid-surface of the shell. Following the work of Lee \textit{et al.}~\cite{lee_geometric_2016}, we assume that each of these imperfections is shaped as a Gaussian dimple,
\begin{align}\label{eqn:geom_gaussiandimple}
w_i(\alpha)= -\delta_i e^{-(\alpha/\alpha_0)^2}, 
\end{align}
where the variable $\alpha$ is the \textit{local} angular distance (spherical coordinate) from the center of the defect, and the constants $\delta_i$ and $\alpha_0$ are, respectively, the amplitude (maximum radial deviation of the mid-surface) and half-angular width of the $i$-th defect. The defect amplitude is non-dimensionalized by the shell thickness, $\overline{\delta}_i=\delta_i/t$. Hereon, when referring to the defect amplitudes (and associated quantities such as the mean and standard deviation of its distribution), we shall mean their dimensionless versions. Also, following a standard in the literature~\cite{kaplan1954nonlinear,koga1969axisymmetric}, we use the following geometric parameter to characterize the defect width:
\begin{align}\label{eqn:lambda}
\lambda=[12(1-\nu^2)]^{1/4}\,\eta^{1/2}\,\alpha_0.
\end{align}

As shown schematically in Fig.~\ref{fig:probdef}(a), the location on the shell's mid-surface of the center of each defect (where the local angle is $\alpha=0$) is defined by the unit radial vector:
\begin{align}\label{eqn:defectposition}
\textbf{e}_{r_i} = \sin{\beta_i}\cos{\theta_i} \textbf{e}_x + \sin{\beta_i}\sin{\theta_i} \textbf{e}_y + \cos{\beta_i} \textbf{e}_z,
\end{align}
where $\beta_i$ and $\theta_i$ are specific values of the \textit{global} zenith (polar) and azimuthal spherical coordinates, $\beta$ and $\theta$, respectively, associated with the $i$-th defect. Note that the \textit{local} angle, $\alpha$ in Eq.~(\ref{eqn:geom_gaussiandimple}), associated with each defect, can be related to $(\beta,\,\theta)$ and their specific values for the defect center $(\beta_i,\,\theta_i)$ but this complicated function is unnecessary to define the geometry. The pole of the shell is located at $\beta=0$ and the origin of the Cartesian coordinate system $(\textbf{e}_x,\,\textbf{e}_y,\,\textbf{e}_z$) is located at the center of the circular base of the hemisphere. The radial distance of the middle surface of the shell is $r_\mathrm{m}(\beta,\,\theta)=R+\sum_{i=1}^Nw_i(\beta,\,\theta)$. The center of each defect is at a distance $b_i=r_\mathrm{m}(\beta_i,\,\theta_i)=R-\delta_i$ from the origin and its radial position vector is $\mathbf{r}_i=b_i\textbf{e}_{r_i}$. The angular separation between two neighbouring defects, $i$ and $j$, whose centers are located at $\textbf{e}_{r_i}$ and $\textbf{e}_{r_j}$, is defined as
\begin{align}\label{eqn:phi_def}
    \varphi_{(i,j)}=\big|\arccos\left(\textbf{e}_{r_i}\cdot \textbf{e}_{r_j}\right) \big|,
\end{align}
with $\varphi_{(i,j)}>90^\circ$ due to the hemispherical geometry. If the $i$-th defect is located at the pole (at $\beta=0$), then $\varphi_{(i,j)}=\beta_j$, but this is not so in the general case of $\beta_i\neq\beta_j\neq0$ and $\theta_i\neq\theta_j$.

First, we will address the buckling problem of a shell containing only two imperfections ($N=2$), seeking to characterize the effect of defect-defect interactions. To this end, with $i=\{1,\,2\}$, we systematically vary the defect amplitudes, $\overline{\delta}_i$, geometric parameter, $\lambda_i$, and their angular (center-to-center) separation, $\varphi_{(1,2)}$. The $i=1$ defect is positioned at the shell pole ($\beta=0$) and the $i=2$ defect at a polar location $\beta_2$, such that $\varphi_{(1,2)}=\beta_2$. We aim to quantify how the knockdown factor of this imperfect shell depends on $\overline{\delta}_i$, $\lambda_i$, and $\varphi_{(1,2)}$.
For the explored $N=2$ configurations, we will consider two cases for the defects' geometric parameters and amplitudes: one with $\lambda=\lambda_1=\lambda_2$ and $\overline{\delta}=\overline{\delta}_1=\overline{\delta}_2$, and another with $\lambda_1\neq\lambda_2$ and $\overline{\delta}_1\neq\overline{\delta}_2$. In both cases, we will find that the knockdown factor, $\kappa$, exhibits a nontrivial behavior as a function of $\varphi_{(1,2)}$; these results are reported in Section~\ref{sec:results_2defects}. The defect-defect interactions will be important when setting up and interpreting the more complex problem of buckling of imperfect shells with a large distribution ($N\gg1$) of defects, introduced next.

In our second and central problem, we build upon past studies on the sensitivity of the buckling of spherical defects containing a \textit{single} defect~\cite{lee_geometric_2016,hutchinson2016buckling,jimenez2017technical,marthelot2017buckling} to now consider imperfect shells with a \textit{large number} ($N\gg1$) of randomly distributed defects. Specifically, we design shells whose defects have a statistical distribution of amplitudes $\overline{\delta}_i$ (lognormally distributed) and positions $\textbf{r}_i$ (distributed according to a random sequential adsorption algorithm). These designs explore a few fixed values of the defect angular width, $\alpha_0$, and hence $\lambda$ through Eq.~(\ref{eqn:lambda}). The probability density function (PDF) of the defect amplitude is:
\begin{align} \label{eqn:lognormal}
    f(\overline{\delta}_i)=\frac{1}{\overline{\delta}_i \sigma \sqrt{2\pi}}
    \exp\left( 
    -\frac{(\ln \overline{\delta}_i -\mu)^2}{2 \sigma^2}
    \right),
\end{align}
where $\mu$ and $\sigma$ are parameters related to the mean defect amplitude
\begin{align} \label{eqn:PDFmean}
    \langle\overline{\delta}\rangle=\exp\left( \mu + \frac{\sigma^2}{2}\right)
\end{align}
and its standard deviation
\begin{align} \label{eqn:PDFstd}
    \Delta\overline{\delta}=\biggl\{  \left[\exp(\sigma^2)- 1 \right]\, \exp\left( 2\mu + \sigma^2\right)\biggl\}^{1/2}.
\end{align}
%
%
%
Note that the logarithms of the defect amplitudes, $\ln(\overline{\delta})$, are normally distributed with mean $\mu$ and standard deviation $\sigma$. One advantage of using a lognormal PDF in seeding the imperfections is that it establishes positive values of $\overline{\delta}_i$, ensuring that we only deal with dimples (and not a combination of dimples and bumps). Also, lognormal distributions of imperfections are used widely in structural reliability analysis~\cite{surendran2005peaking, low2013new, le2011unified, le2020failure}.

The position of each defect, defined in Eq.~(\ref{eqn:defectposition}), is seeded randomly onto the shell, one by one, using a random sequential adsorption algorithm~\cite{lopez-pamies_nonlinear_2013, shrimali_simple_2019, lefevre_nonlinear_2017}, which is commonly used to generate nearly isotropic porous structures. Here, we modify the volumetric case to randomly distribute circles on the hemispherical surface of a shell. Overlaps are avoided by setting a minimum angular separation between defects, $\varphi_\textrm{min}$. The seeding procedure stops when the spherical cap delimited by a maximum zenith angle, $\beta_\textrm{max}$, no longer admits a new defect. We chose $\beta_\textrm{max}=60^\circ$ to minimize interaction effects with the boundaries. For example, this seeding procedure with $\lambda=1$ and $\beta_\mathrm{max}=60^\circ$ yields $N\approx80$ defects. The seeding algorithm will be described in more detail in~Section~\ref{sec:methods_experiments_fabrication}.

The hemispherical, imperfect shell with $N$ random defects, is clamped at its equator and loaded under negative pressure, $p_0$. When the imposed pressure difference reaches the critical value, $p_\mathrm{max}$, the shell buckles through a subcritical bifurcation. The case of a single imperfection, $N=1$, has been addressed in previous studies~\cite{lee_geometric_2016,hutchinson2016buckling,jimenez2017technical,marthelot2017buckling, abbasi_probing_2021}, as described in the introduction. In the present work, we seek to characterize the knockdown factor of shells with $N\geq 2$. First, we focus on the defect-defect interactions in the $N=2$ case, and then, we turn to the probabilistic case with $N\gg1$ defects. We systematically vary the mean, $\langle\overline{\delta}\rangle$, and standard deviation, $\Delta\overline{\delta}$, of the seeding lognormal distribution, while fixing $\lambda$ and $\varphi_\textrm{min}$. Statistical ensembles of imperfect shells with statistically equivalent configurations are produced for each set of parameters, $(\langle\overline{\delta}\rangle,\,\Delta\overline{\delta},\,\lambda,\,\varphi_\textrm{min})$. 

Producing a large number of realizations is impractical in experiments. As such, in the second stage of our investigation, we will perform a systematic statistical investigation using the FEM simulations exclusively. Trust on the FEM will be built up by a prior direct quantitative validation against a few specific experimental cases. For practical reasons, the experiments (and, hence, the FEM validation) will be performed for shells with \textit{outward} (bump) defects, instead of the \textit{inward} (dimple) defects in Eq.~(\ref{eqn:geom_gaussiandimple}). This choice is motivated by limitations during the fabrication of the experimental specimens, as detailed in Section~\ref{sec:methods_experiments_fabrication}.

The experimentally validated FEM will then be leveraged to simulate the buckling of statistical ensembles of imperfect shells, designed with either $N=2$ or $N\gg1$ defects, as described above. The probabilistic results will be interpreted in light of previous findings~\cite{lee_geometric_2016} for $N=1$, together with the case of $N=2$ (Section~\ref{sec:results_2defects}). The latter includes the possibility of defect-defect interactions, which are important for the $N\gg1$ probabilistic case. Ultimately, the primary question we tackle is: \textit{Given an input set of statistics for the design geometry of the imperfect shells, what are the output statistics of the resulting knockdown factors, $\kappa$, as characterized by the probability density function, $f(\kappa)$?}

\section{Methods -- Experiments}
\label{sec:methods_experiments}

We proceed by describing the experimental fabrication and characterization of shell specimens containing multiple defects, as well as the experimental protocol followed to measure their critical buckling conditions under pressure loading. 

Previous experimental studies on shell buckling~\cite{lee_geometric_2016, marthelot2017buckling, yan_buckling_2020, yan2021magneto, abbasi_probing_2021} have employed a coating technique~\cite{lee_fabrication_2016} to fabricate elastomeric shells of nearly constant thickness, containing a single well-defined geometric imperfections. Toward tackling the problem defined in Section~\ref{sec:problem_definition}, while building on past work, we have developed a novel experimental technique using 3D-printed molds, enabling the robust fabrication of shells containing \textit{multiple} defects, whose geometry, number, and layout can be designed precisely. 

The exact geometry of the experimental shells is different from the idealized geometry described in Section~\ref{sec:problem_definition} (cf. Fig.~\ref{fig:probdef}); however, the high precision buckling tests described in Section~\ref{sec:methods_experiments_criticalpressure} will serve as a thorough quantitative validation of FEM (Section~\ref{sec:validation}). For validation of the simulations, the full geometry of the fabricated shells is quantified using X-Ray micro-computed tomography ($\mu$CT) and imported into the FEM (Section~\ref{sec:methods_FEM}). After validating the FEM, the buckling of imperfect shells with $N \geq 2$ is more systematically explored using the simplified geometry of Section~\ref{sec:problem_definition}; these results will be presented in Sections~\ref{sec:results_2defects} and~\ref{sec:results_distribution}.

\subsection{Design and fabrication of imperfect shell specimens}
\label{sec:methods_experiments_fabrication}

\begin{figure}[h]
    \centering\includegraphics[width=4in]{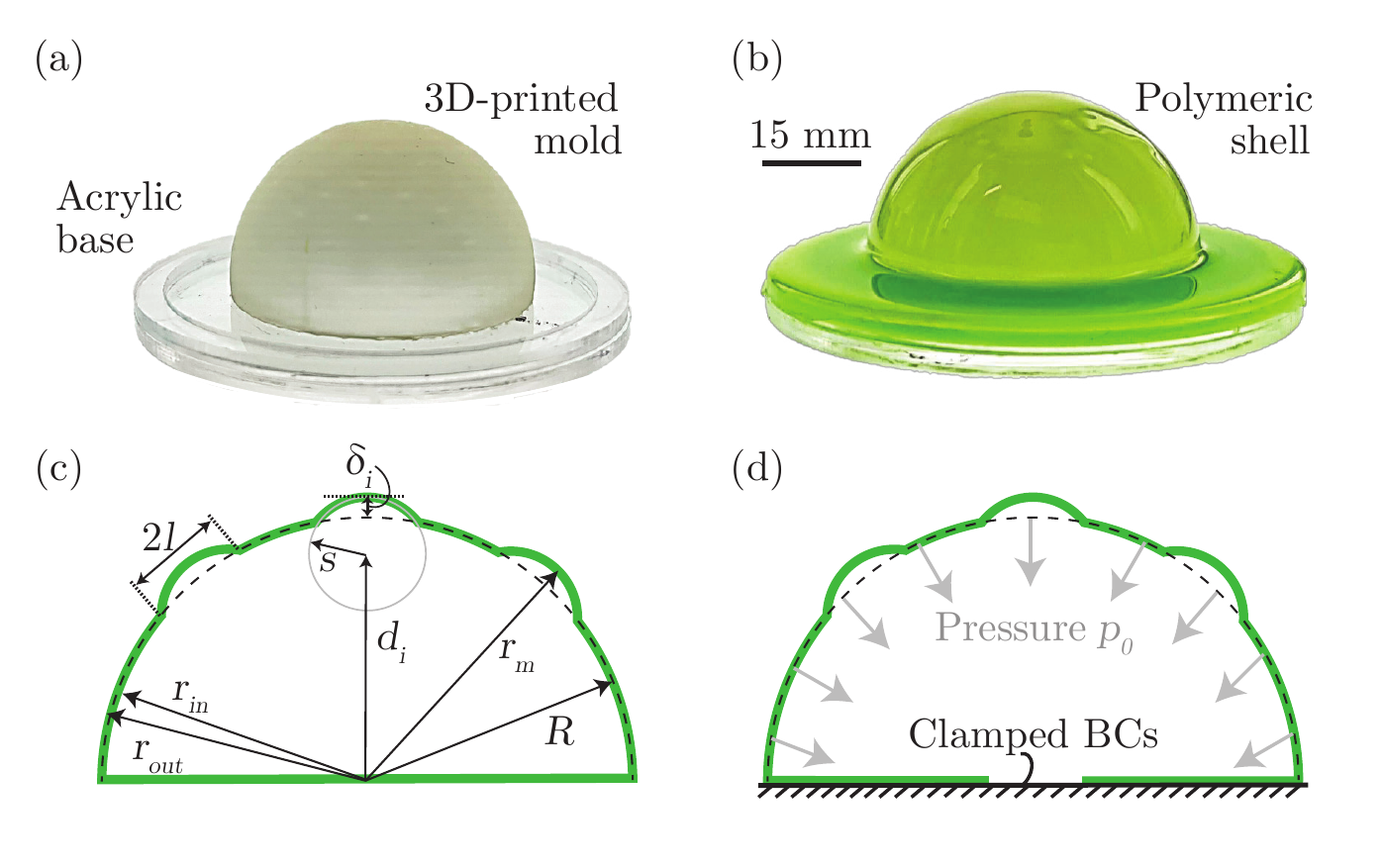}
    \caption{Fabrication and design of the imperfect shell specimens. (a) Representative photograph of the 3D-printed mold containing several defects (bumps). Due to the small size of the bumps, they are not visible at naked eye in the photograph. (b) The mold is coated with a liquid film of VPS-32, which upon curing and demolding yields the thin hemispherical shell. (c) Schematic diagram of she shell geometry.
    (d) The hemispherical shell is clamped at the equator and depressurized to measure the critical buckling conditions.
    }
    \label{fig:fabrication}
    \end{figure}

We produced textured hemispherical molds using a desktop stereolithography 3D-printer (Form 2 Formlabs) using  Clear V4 resin (see Fig.~\ref{fig:fabrication}a), with a resolution of $25\,\mu$m per layer. Throughout, the nominal radius of the molds is fixed at $R = 25.4\,$mm. 

The imperfections are introduced by design into the surface of the spherical mold, with several small spherical protrusions of radius $s\ll R$ to produce defects as \textit{bumps} (see Fig.~\ref{fig:fabrication}c). The center of each of these spherical bumps is located at a distance $d_i$ from the center of the nominal sphere of radius $R$, so that they produce a maximum radial distance $\delta_i = d_i+s-R$. The defect width is defined by the intersection between the bump and the nominal sphere of the mold, and computed from geometry~\cite{Weisstein} as
\begin{align}\label{eqn:defect_width}
l = \frac{1}{2d_i}\sqrt{4d_i^2R^2 - (d_i^2 - s^2 + R^2)^2}.
\end{align}

A number of bumps protruding from a hemisphere is seeded onto the mold surface, with well-defined statistics for $\overline{\delta}_i$ and $l$, and using the same random sequential adsorption algorithm mentioned in Section~\ref{sec:problem_definition}. The seeding is performed within a spherical cap of the mold, delimited by a threshold value of the zenith angle $\beta_\mathrm{max}$, to avoid potential interactions with the equatorial boundary when the shell is fabricated (more on this below). Within this cap, we randomly sample the angular position of the center of each $i$-th defect, 
\begin{equation}
\Bigl(\beta_i,\theta_i\Bigl)=\Bigl(\arccos\bigl({1-x_{\beta}(1-\cos{\beta_\mathrm{max}})\bigl)},\, 2\pi x_{\theta}\Bigl),
\end{equation}
where $x_{\beta}\in[0,\,1]$ and $x_{\theta}\in[0,\,1]$ are two random variables with an equal probability, to locate the defect anywhere on the hemisphere cap. The algorithm also imposes a constraint on the angle between the centers of any pair of defects (with indices $i$ and $j$),
\begin{align}\label{eqn:phi_min}
\varphi_{(i,j)} \geq \varphi_\mathrm{min},
\end{align}
such that $\varphi_\textrm{min}\geq 2\alpha_0 = 2\arcsin{\left(l/R\right)}$.
%
%
For each shell design with fixed $l$, every new $i$-th defect is seeded at a random location of the shell and tested for Eq.~(\ref{eqn:phi_min}), with respect to all other already existing defects ($j=1,\ldots, i-1$). If Eq.~(\ref{eqn:phi_min}) is satisfied, a new $i+1$ defect is seeded with the same procedure. If Eq.~(\ref{eqn:phi_min}) is not satisfied, this $i$-th defect is removed, and a new random location is generated until a valid condition is attained, up to a given maximum number ($10^6$) of attempts. When this number of attempts is reached, the seeding process stops, setting the final number of defects, $N$.

The amplitude of each of the defects is treated as a random variable using a normal distribution, $\overline{\delta}_i\sim \mathrm{Normal}(\langle\overline{\delta}\rangle,\, \Delta\overline{\delta}^2)$, with mean $\langle\overline{\delta}\rangle$ and standard deviation $\Delta \overline{\delta}$. Note that the normal distribution used for the experimental specimens (with \textit{outward} bumps) is only employed in the experiments geared to validate the FEM simulations in Section~\ref{sec:validation}. The FEM simulations in Sections~\ref{sec:results_2defects} and~\ref{sec:results_distribution} use designs generated with a lognormal distribution (with \textit{inward} dimple); cf. Eqs.~(\ref{eqn:geom_gaussiandimple}) and~(\ref{eqn:lognormal}). The reason for this disparity is practical: we first performed the experiments before realizing that lognormal-distributed defects were a better choice, but it was then impractical to redo all experiments with the new design. Regardless of the design details, the experiments are appropriate for the detailed validation of the FEM before turning to the more thorough and systematic numerical exploration of the problem defined in Section~\ref{sec:problem_definition}. 

We have designed and 3D-printed eight different molds, whose geometric parameters are detailed in Table~\ref{table:shell_specimens}. Four of the molds had bumps in the spherical cap within $\beta_\mathrm{max} = 20^\circ$ and the other four within $\beta_\mathrm{max} = 60^\circ$. The explored ranges of the design parameters were: $0.5 < \lambda < 4.0$, $0.5 < \langle \overline{\delta}\rangle < 2.5$, and $5^\circ < \varphi_\textrm{min} < 35^\circ$. Consequently, these parameters set the range of the number of defects: $1 \leq N \leq 30$. Note that Shell 4 corresponds to a mold containing a single defect, which is achieved by setting $\varphi_{\mathrm{min}}=\beta_{\mathrm{max}}$.

 \begin{table}[h!]
    \caption{Design parameters of the shells with bumps fabricated in the experiments (see Section~\ref{sec:methods_experiments_fabrication}) and simulated via FEM for validation (see Section~\ref{sec:validation}): average $\langle\overline{\delta}\rangle$ and standard deviation $\Delta \overline{\delta}$ of defect amplitude, geometric (width) parameter $\lambda$,  minimum angle between defect centers $\varphi_\mathrm{min}$, maximum zenith angle $\beta_\mathrm{max}$, and number of defects $N$.}
    \label{table:shell_specimens}
    \centering
    \small
    \renewcommand{\arraystretch}{1}
    \begin{tabular}{c c c c c c c}
    \hline\hline
    \multicolumn{1}{l}{Specimen} &
    \multicolumn{1}{c}{$\lambda$} &
    \multicolumn{1}{c}{$\langle \overline{\delta}\rangle$} &
    \multicolumn{1}{c}{$\Delta \overline{\delta}$} &
    \multicolumn{1}{c}{$\varphi_\textrm{min}$} &
    \multicolumn{1}{c}{$\beta_\textrm{max}$} &
    \multicolumn{1}{c}{$N$} \\
        \multicolumn{1}{c}{No.} &
    \multicolumn{1}{c}{[-]} &
    \multicolumn{1}{c}{[-]} &
    \multicolumn{1}{c}{[-]} &
    \multicolumn{1}{c}{[$^\circ$]} &   \multicolumn{1}{c}{[$^\circ$]} &
    \multicolumn{1}{c}{[-]} \\
    \hline\hline
    \multicolumn{1}{c}{Shell 1} & \multicolumn{1}{c}{$0.539$} & \multicolumn{1}{c}{$0.643$}  & \multicolumn{1}{c}{$0.000$} & \multicolumn{1}{c}{$6.189$} & \multicolumn{1}{c}{$20$} & \multicolumn{1}{c}{$7$}\\
    \multicolumn{1}{c}{Shell 2} & \multicolumn{1}{c}{$1.056$} & \multicolumn{1}{c}{$0.575$} & \multicolumn{1}{c}{$0.000$} & \multicolumn{1}{c}{$9.446$} & \multicolumn{1}{c}{$20$} & \multicolumn{1}{c}{$7$}\\
    \multicolumn{1}{c}{Shell 3} & \multicolumn{1}{c}{$0.553$} & \multicolumn{1}{c}{$0.794$} & \multicolumn{1}{c}{$0.180$} & \multicolumn{1}{c}{$4.790$} & \multicolumn{1}{c}{$20$} & \multicolumn{1}{c}{$24$}\\
    \multicolumn{1}{c}{Shell 4} & 
    \multicolumn{1}{c}{$2.205$} & \multicolumn{1}{c}{$0.769$} & \multicolumn{1}{c}{$0.000$} & 
    \multicolumn{1}{c}{$20.000$} & \multicolumn{1}{c}{$20$} & \multicolumn{1}{c}{$1$}\\
    \multicolumn{1}{c}{Shell 5} & 
    \multicolumn{1}{c}{$1.571$} & \multicolumn{1}{c}{$0.698$} & \multicolumn{1}{c}{$0.000$} & 
    \multicolumn{1}{c}{$14.787$} & \multicolumn{1}{c}{$60$} & \multicolumn{1}{c}{$30$}\\
    \multicolumn{1}{c}{Shell 6} &
    \multicolumn{1}{c}{$2.188$} & \multicolumn{1}{c}{$0.757$} & \multicolumn{1}{c}{$0.166$} & 
    \multicolumn{1}{c}{$18.717$} & \multicolumn{1}{c}{$60$} & \multicolumn{1}{c}{$17$}\\
    \multicolumn{1}{c}{Shell 7} & 
    \multicolumn{1}{c}{$1.889$} & \multicolumn{1}{c}{$0.865$} & \multicolumn{1}{c}{$0.000$} & 
    \multicolumn{1}{c}{$18.488$} & \multicolumn{1}{c}{$60$} & \multicolumn{1}{c}{$20$}\\
    \multicolumn{1}{c}{Shell 8} & 
    \multicolumn{1}{c}{$3.519$} & \multicolumn{1}{c}{$2.143$} & \multicolumn{1}{c}{$0.000$} & 
    \multicolumn{1}{c}{$34.954$} & \multicolumn{1}{c}{$60$} & \multicolumn{1}{c}{$5$}\\
    \hline\hline
    \end{tabular}
    \normalsize
    \end{table}
    
Each hemispherical shell is fabricated by pouring vinylpolysiloxane (VPS-32, Elite Double 32, Zhermack) polymer onto the respective 3D-printed mold, as shown in Fig.~\ref{fig:fabrication}(b). To emulate clamped boundary conditions of the fabricated shells during testing, a distance of $3\,$mm is set between the lower part of the printed mold and the center of the hemisphere, allowing for the formation of a thick polymeric equatorial lip during the shell fabrication (see Fig.~\ref{fig:fabrication}b).
The mold was mounted concentrically on the circular recess of a flat plate (depth 3\,mm), whose top surface was aligned with the equator of the mold. Upon pouring, this recess was filled with the VPS-32 polymer solution, while avoiding overflow, to form an equatorial lip. The thickness of this lip (3\, mm) was significantly thicker than the typical values of the shell thickness ($t\approx 0.3\,$mm for Shells 1-7 and $t\approx 0.4\,$mm for Shell 8). As such, this lip served to emulate clamped boundary conditions during the buckling tests (more on these below). 

The VPS-32 polymer solution was prepared by mixing a base and a curing agent (1-1 weight ratio) in a centrifugal mixer (ARE-250, Thinky USA Inc., Laguna Hills, CA) for $200\,$rpm/sec clockwise and $220\,$rpm/sec counterclockwise, a process that also removes air bubbles. After mixing, the VPS-32 solution was only poured onto the molds after a quiescent waiting time of 3\,min to control the viscosity of the solution when pouring, setting an appropriate value of the thickness of the gravity-driven lubrication flow down the surface of the mold~\cite{lee_fabrication_2016}. After a curing time of 20\,min, the liquid VPS-32 film solidified, and the resulting elastic shell containing a distribution of defects was peeled from the mold and tested under internal pressure, $p_0$ (Fig.~\ref{fig:fabrication}d). 

Note that the bumps protruding from the spherical mold act as topographic barriers to the gravity-driven lubrication flow of VPS-32 prior to curing, causing local modifications in both the geometry of the mid-surface and thickness of the otherwise nearly constant film thickness on a spherical substrate. These specifics of the lubrication flow during fabrication are the reason why, in the experiments, we decided to consider imperfections as \textit{outward} bumps instead of the \textit{inward} dimples mentioned in Section~\ref{sec:problem_definition}. Had we used dimpled (instead of bumpy) molds,  each of the topographic depressions would have acted as a fluid-accumulating basin causing significant increases of the film thickness there. In that case, the shape of the resulting individual defects would  be undesirably shaped very differently than the target dimpled profile of Eq.~(\ref{eqn:geom_gaussiandimple}). These
fabrication limitations underlie our choice for experimental designs with bumpy defects, which are primarily geometric, albeit still with some degree of thickness variation.

\subsection{Characterization of the geometry of the experimental specimens}
\label{sec:methods_experiments_characterization}

\begin{figure}[h]
    \centering\includegraphics[width=\textwidth]{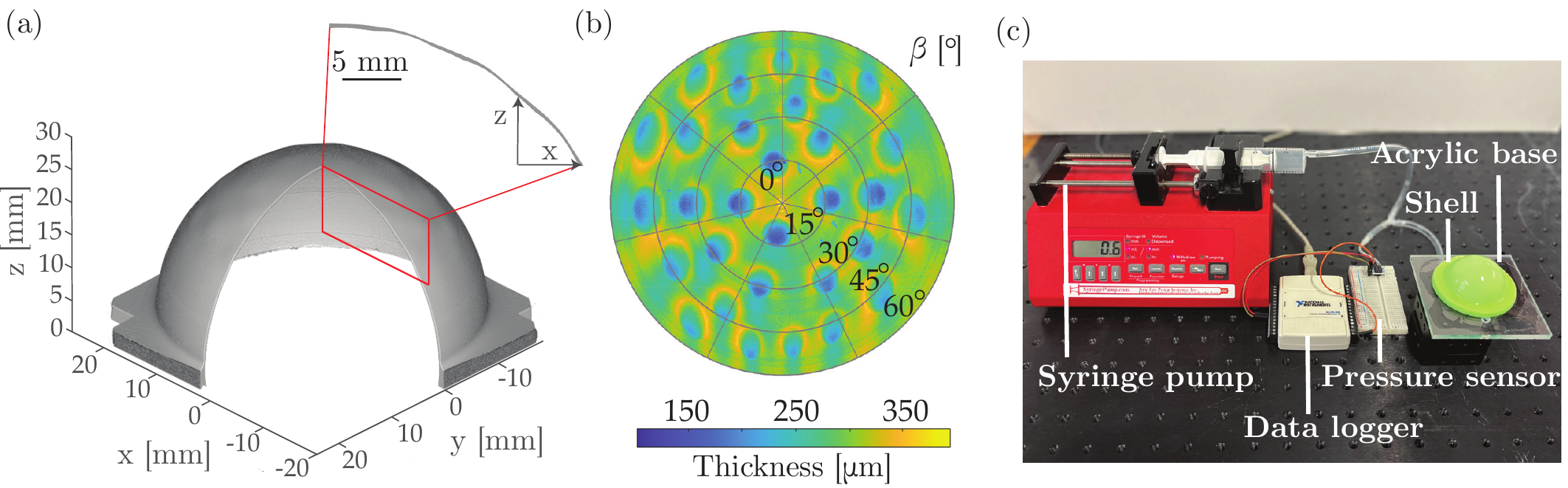}
    \caption{Geometric characterization and apparatus used to perform buckling tests of the shell specimens. (a) Reconstructed 3D image of a shell obtained using X-ray micro-computed tomography ($\mu$CT). Shell 5 (see parameters in Table~\ref{table:shell_specimens}) is used as a representative example. Inset: Magnification of a cross-sectional cut in the $x$-$z$ plane. (b) Thickness profile, $t(\beta,\,\theta)$, of the shell shown in (a) obtained after post-processing of the $\mu$CT data. (c) Experimental apparatus  used to measure the buckling pressure of the hemispherical shell specimens. 
    }
    \label{fig:microct}
    \end{figure}

For the detailed validation of FEM against experiments, we will require knowledge of the full geometry of the imperfect shell specimens listed in Table~\ref{table:shell_specimens}, which we characterized using X-ray micro-computed tomography ($\mu$CT). Our equipment ($\mu$CT100, Scanco Medical) offers a scanning resolution of 25.4\,$\mu$m (voxel size). Each specimen is positioned on a 360-degree rotary stage of the $\mu$CT, which, after image processing of a 2D stack of images, yields a volumetric (3D) reconstruction of the shell using ImageJ~\cite{schindelin2012fiji}. A representative reconstruction of such a 3D image is presented in Fig.~\ref{fig:microct}(a), for Shell 5 (cf. Table~\ref{table:shell_specimens}). In the inset of Fig.~\ref{fig:microct}(a), we present an arbitrary cross-sectional cut of the reconstruction in the $x$-$z$ plane, exhibiting the thickness profile of the imperfect shell.

An in-house Matlab~\cite{MATLAB:2021a} algorithm was then employed to post-process the 3D images and obtain several geometric quantities. Specifically, we perform edge detection on the $\mu$CT images to extract the inner and outer surfaces of the shell in spherical coordinates $(r,\,\beta,\,\theta)$. These surfaces are represented, respectively, by their inner, $r_\mathrm{in}(\beta,\,\theta)$, and outer, $r_\mathrm{out}(\beta,\,\theta)$, radial positions, from which we then compute the radial position of the shell's mid-surface, $r_\mathrm{m}(\beta,\,\theta)=[r_\mathrm{out}(\beta,\theta)+r_\mathrm{in}(\beta,\theta)]/2$, and the shell-thickness profile $t(\beta,\,\theta)=r_\mathrm{out}(\beta,\theta)-r_\mathrm{in}(\beta,\theta)$. The edge-detection algorithm requires a segmentation threshold, which is sensitive for the correct determination of $t$. We calibrate this threshold value using a digital microscope (VHX, Keyence) to independently measure the shell thickness of a $2\times10\,\mathrm{mm}^2$ portion of a sacrificial specimen that is cut to expose the cross-section. The segmentation threshold is systematically adjusted until the value of $t$ measured by microscopy on this sacrificial sample matches that obtained by image processing of the $\mu$CT images on the same sample.

In Fig.~\ref{fig:microct}(b), we show a representative measurement of the thickness profile, $t(\beta,\,\theta)$,
for Shell 5. This thickness profile can then be used to quantify the geometry of the imperfections. Due to the specifics of the fabrication procedure described in Section~\ref{sec:methods_experiments_fabrication}, each defect comprises a combination of a geometric imperfection (of the mid-surface, away from a perfect hemisphere) combined with a small degree of a through-thickness imperfection. Defects produced by small bumps ($s<1\,$mm) in the mold had a nearly axisymmetric thickness about their center. By contrast, larger defects had asymmetric thicknesses due to VPS-32 polymer accumulating upstream of the bump in the gravity-driven lubrication flow during coating. The full geometric profiles of each of the fabricated shells listed in Table~\ref{table:shell_specimens} was characterized thoroughly using the $\mu$CT. Further characterizing individual defects by decoupling the geometric and through-thickness imperfections is beyond the scope of the present work. Instead, the full 3D geometry characterized experimentally was imported into the FEM simulation, allowing for a direct and accurate comparison of the buckling pressures between physical and numerical experiments for validation purposes (see Section~\ref{sec:validation}). 

Eventually, we will report experimental results for the knockdown factor, defined in Eq.~(\ref{eqn:knockdown_def}) as $\kappa=p_\mathrm{max}/p_\mathrm{c}$, for all of the fabricated specimens. The protocol to measure $p_\mathrm{max}$ will be detailed below, in Section~\ref{sec:methods_experiments_criticalpressure}. Still, to compute $\kappa$, we will also need $p_\mathrm{c}$, the classic theoretical prediction for a perfect shell defined in Eq.~(\ref{eqn:zolley}) and, consequently, a nominal value of the thickness, $t_\mathrm{0}$, for the case without defects. For this purpose, we fabricated five nearly perfect (no defects) shells using a stainless steel sphere as the mold ($R = 25.4\,$mm, same as the nominal radius of the 3D-printed molds). The conditions for the preparation of the VPS-32 solution, pouring, and curing were identical to the case of imperfect shells described above. A spherical cap was cut from these nearly perfect shells at $\beta=45^\circ$ and the thickness was measured using the digital microscope to be $t_\mathrm{0}=300 \pm 10\,\mu$m ($\eta = R/t_0 =85$).

\subsection{Measurement of the critical buckling pressure}
\label{sec:methods_experiments_criticalpressure}

After fabrication and $\mu$CT scanning, we measured the critical buckling pressure of each of the imperfect shell specimens listed in Table~\ref{table:shell_specimens}, using an experimental protocol identical to that developed in our previous work~\cite{lee_geometric_2016,marthelot2017buckling,abbasi_probing_2021,yan2021magneto}. For completeness, we summarize this protocol next. 

In Fig.~\ref{fig:microct}(c), we show a photograph of the experimental apparatus. A shell specimen was mounted onto an acrylic plate, sealing its equator with a thin layer of VPS-32 polymer. The thick equatorial lip of the shell, together with this layer of VPS-32, enforced clamped boundary conditions at the equator. An additional thin water film was deposited at the joint between the shell equator and the acrylic plate to achieve airtightness. The center of the acrylic-plate mount contains a through-hole and is connected to a syringe pump (NE-300, New Era Pump Systems, Inc), a pressure sensor (HSCDRRN005NDAA5, Honeywell Sensing and Productivity Solutions), which is itself linked to a data logger (NI USB-6009, National Instruments). This system is used to depressurize the shell specimen by withdrawing air using the syringe pump (at the constant rate of $0.6\,$mL/min), while recording the internal-pressure signal using the pressure sensor and the data logger. The gradual increase of the pressure differential between the inside of the shell and the outside (at atmospheric pressure) eventually causes the shell to buckle at the critical value, $p_\mathrm{max}$, past which the measured pressure drops sharply. 

Each shell specimen is tested ten times for quantification of the experimental uncertainties, and the average value is reported. The experimental measurements of $p_\mathrm{max}$ for each of the specimens listed in Table~\ref{table:shell_specimens} will be reported in Section~\ref{sec:validation} (Fig.~\ref{fig:validation}b).

\section{Methods -- Finite element modeling}
\label{sec:methods_FEM}

In parallel to the experiments, we performed FEM simulations using the commercial package ABAQUS/Standard~\cite{ABAQUS:2014}.
In previous works for shells with a single defect~\cite{lee_geometric_2016,yan_buckling_2020,yan2021magneto,abbasi_probing_2021}, symmetries were exploited to reduce computational costs. By contrast, the spatial distribution of imperfections in the present study requires the numerics to be tackled as fully 3D. We used S4R shell elements with reduced integration points, allowing for finite membrane strains. The hemispherical shell is divided into four quarters, each composed of 150 elements in both the meridional and azimuthal directions. This discretization choice was deemed appropriate after a mesh-convergence study and will be supported further by the successful validation against experiments (Section~\ref{sec:validation}). We employed a static Riks solver, selecting an initial arc length of increment of $10^{-1}$, with minimum and maximum increment sizes of $10^{-5}$ and $0.5$, respectively. Throughout, geometric nonlinearities were considered. The VPS-32 elastomer was modeled as a neo-Hookean solid with a Poisson's ratio of $\nu=0.5$ (assuming incompressibility) and a Young's modulus of $E = 1.25 \pm 0.01\,$MPa (the measured experimental value).

We recall that the FEM simulations were tackled in two stages. First, we performed a direct validation against experiments (using the experimentally measured geometries); these validation results are provided in Section~\ref{sec:validation}. Then, having built trust on the simulations upon their successful validation, the FEM was employed for the thorough investigation of the problem defined in Section~\ref{sec:problem_definition} using more idealized geometries; these results are provided in Sections~\ref{sec:results_2defects} and~\ref{sec:results_distribution}.

For the validation-purposed simulations, the 3D geometric models were imported from the $\mu$CT tomographic scans of the eight experimental specimens listed in Table~\ref{table:shell_specimens}. From these scans, as described in Section~\ref{sec:methods_experiments_characterization}, we extracted the hemispherical profiles of the outer radius, $r_\mathrm{out}$, inner radius, $r_\mathrm{in}$, and thickness, $t$; all functions of the voxelated spherical coordinates $(\beta,\,\theta)$. In ABAQUS, it is difficult to generate a mesh directly from these raw $\mu$CT data (point clouds). Alternatively, we first meshed a perfect hemispherical shell, where each node was assigned angular positions and a radius. Nodal displacements were imposed on the original hemispherical mesh such that they matched the interpolated from the $\mu$CT values of $r_\mathrm{out}$ for the same angular coordinates. A nodal thickness equal to the interpolated value from the $\mu$CT $t(\beta,\,\theta)$ profile was then applied at each note. Finally, an offset was imposed using the keyword \texttt{*OFFSET} to inform ABAQUS that the specified mesh corresponds to the outer surface of the shell elements.

The FEM simulations of the imperfect shells containing Gaussian dimples were done analogously to the validation simulations described above, but with the idealized geometry. First, we created a perfect hemispherical mesh, whose nodes were then displaced according to the desired design set by the parameters $(\langle\overline{\delta}\rangle,\,\Delta\overline{\delta},\,\lambda,\,\varphi_\textrm{min})$, with each defect shaped according to Eq.~(\ref{eqn:geom_gaussiandimple}). In this case, since the topography of imperfections is specified at the middle surface of the shell, no offset needs to be imposed on the reference surface mesh. In all of these simulations of dimpled shells with $N\geq2$, a constant nodal thickness is applied to each node, which was fixed such that $\eta=110$. 

\section{Validation of the FEM simulations}
\label{sec:validation}

To gain confidence in the high-fidelity of our 3D FEM simulations, we followed two stages. First, we performed a verification against published results for the one-defect case. Then, we validated the FEM simulations against the experiments described in Section~\ref{sec:methods_experiments}.

The first verification stage against existing results is important because the previous studies in Refs.~\cite{lee_geometric_2016,yan_buckling_2020,yan2021magneto}, for shells containing only one defect at the pole, made use of circular symmetry (about the vertical axis). Hence, these previous simulations simplified the computational domain to be axisymmetric (2D) to reduce computational cost. By contrast, the present simulations were designed for the many-defects problem, using 3D shell elements, and involving a fully 3D non-axisymmetric geometry. For the case of one defect located at the pole ($\beta=0$), we compared the present 3D FEM to the previous 2D FEM results~\cite{lee_geometric_2016}, for different values of $\overline{\delta}$ and $\lambda$, obtaining, not surprisingly, excellent quantitative agreement between the two. Additional simulations were performed for different polar locations of the defect (while also varying $\overline{\delta}$ and $\lambda$), finding that $\kappa$ remained unmodified as long as the defect was located within $0\leq\beta\leq 60^\circ$; closer to the equator, interactions with the clamped boundary become important. 
We recall that in Ref.~\cite{lee_geometric_2016}, with the defect at the pole, boundary effects were deemed negligible by comparing the FEM results for a hemispherical shell with shell-theory predictions for a complete spherical shell. Shells with the even more restrictive value of $\beta_\mathrm{max} = 20^\circ$ will be considered below (cf. Table~\ref{table:shell_specimens}), but that choice will also present some issues. Therefore, the value of $\beta_\mathrm{max}=60^\circ$ was kept constant through the subsequent simulations of shells with two defects (Section~\ref{sec:results_2defects}) and with random distributions of defects (Section~\ref{sec:results_distribution}).

\begin{figure}[!h]
    \centering\includegraphics[width=\textwidth]{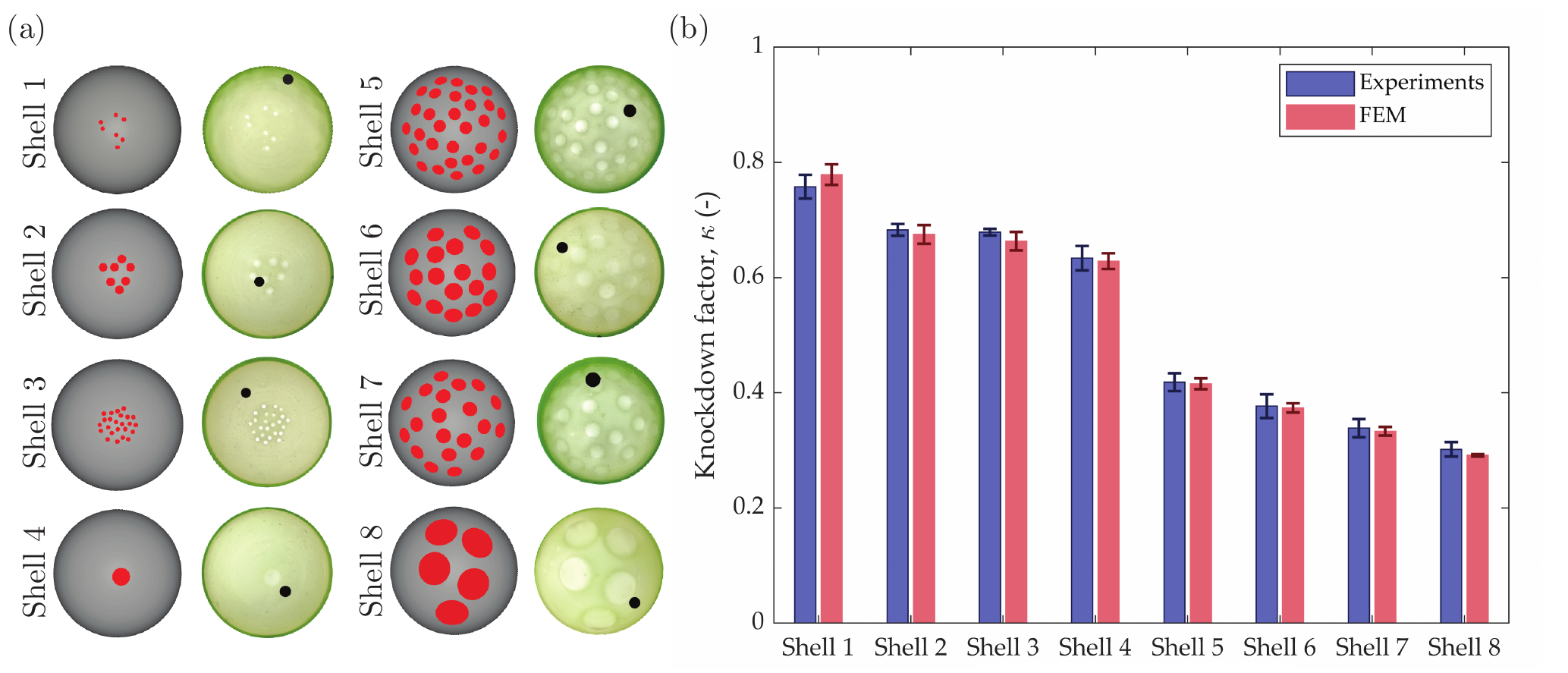}
    \caption{Validation of the FEM simulations against experiments. (a) Top views of each of the imperfect shells, showing both the design drawing (left columns) and the corresponding photograph of the fabricated shells (right columns). On the photographs, a black marker indicates the location where buckling occurs immediately past $p_\mathrm{max}$. (b) Plot of the knockdown factor, $\kappa$, for each of the shells tested in the experiments (blue bars) and computed in FEM (red bars). Additional information on the error bars is provided in the text. In both (a) and (b), the labels -- ``Shell 1'' \ldots ``Shell 8'' -- correspond to Table~\ref{table:shell_specimens}.
    }
    \label{fig:validation}
    \end{figure}

Having built up trust in the 3D FEM for the one-defect case, we now perform a validation against experiments by comparing the simulated and measured values of the knockdown factor, $\kappa$, for each of the eight shells listed in Table~\ref{table:shell_specimens}. In Fig.~\ref{fig:validation}(a), we present top views of the imperfect shells, showing both the design drawing (left columns, where we highlight the location of the defects), and the corresponding photographs (right columns) of the fabricated shells. In Fig.~\ref{fig:validation}(b), we plot the experimentally measured and FEM-computed values of $\kappa$ (blue and red bars, respectively) for each of the shells. The error bars in the experimental results correspond to the standard deviation of 10 tests, and those for FEM correspond to the propagation of errors originating from the experimental uncertainty in the Young's modulus of the VPS-32 elastomer ($E = 1.25 \pm 0.01\,$MPa). The typical uncertainties of $\kappa$ ($\approx 3\%$ for experiments and $\approx 2\%$ for FEM) convey the high-precision of our framework. At this stage, the specimens are representative choices,  exploring broadly the wide range of possible parameters and no structure should be inferred from the data. Therefore, the identification labels of each shell were simply ordered with decreasing $\kappa$. A systematic investigation of parameters will follow in Sections~\ref{sec:results_2defects} and ~\ref{sec:results_distribution}. What is important to appreciate in Fig.~\ref{fig:validation}(b) is the remarkable agreement between experiments and FEM, within a few percent; the discrepancies are at most 3.4\% (for Shell 8) and as low as 0.7\% (for Shell 5), and more typically $\approx$1-2\% for the other shells. 

In each of the photographs in Fig.~\ref{fig:validation}(a), we have also manually marked (with a black dot) the locus of the buckling event observed immediately past $p_\mathrm{max}$. In general, shells containing a distribution of defects with $\lambda<1$ and delimited by $\beta_\mathrm{max} = 20^\circ$ (\textit{e.g.}, Shells 1 and 3) tend to buckle away from the location of the bumpy defects. By contrast, shells with $\lambda>1$ and delimited by $\beta_\mathrm{max} = 60^\circ$ (\textit{e.g.}, Shells 6 and 8) buckle within, or in the proximity of, the designed region of imperfections. The fact that the values of $\kappa$ are in excellent agreement between experiments and FEM conveys that this qualitative difference is not due to experimental artifacts induced during fabrication. Still, to ensure that the buckling capacity is dictated by the designed regions of imperfections (within $\beta_\mathrm{max}$), and not by other subtler effects, we set $\beta_\mathrm{max} = 60^\circ$. Further rationalizing the buckling location goes beyond the scope of the present work.

\section{Buckling of imperfect shells containing two defects}
\label{sec:results_2defects}

Having established our methodology and validated the FEM against experiments, we proceed by focusing on the simulations to study the buckling of shells containing only two defects. We investigate the effect of defect-defect interactions in dictating the knockdown factor, compared to the one-defect case. These results will be important in Section~\ref{sec:results_distribution}
when interpreting the more complex case of shells containing many imperfections.

\begin{figure}[!h]
    \centering\includegraphics[width=0.88\textwidth]{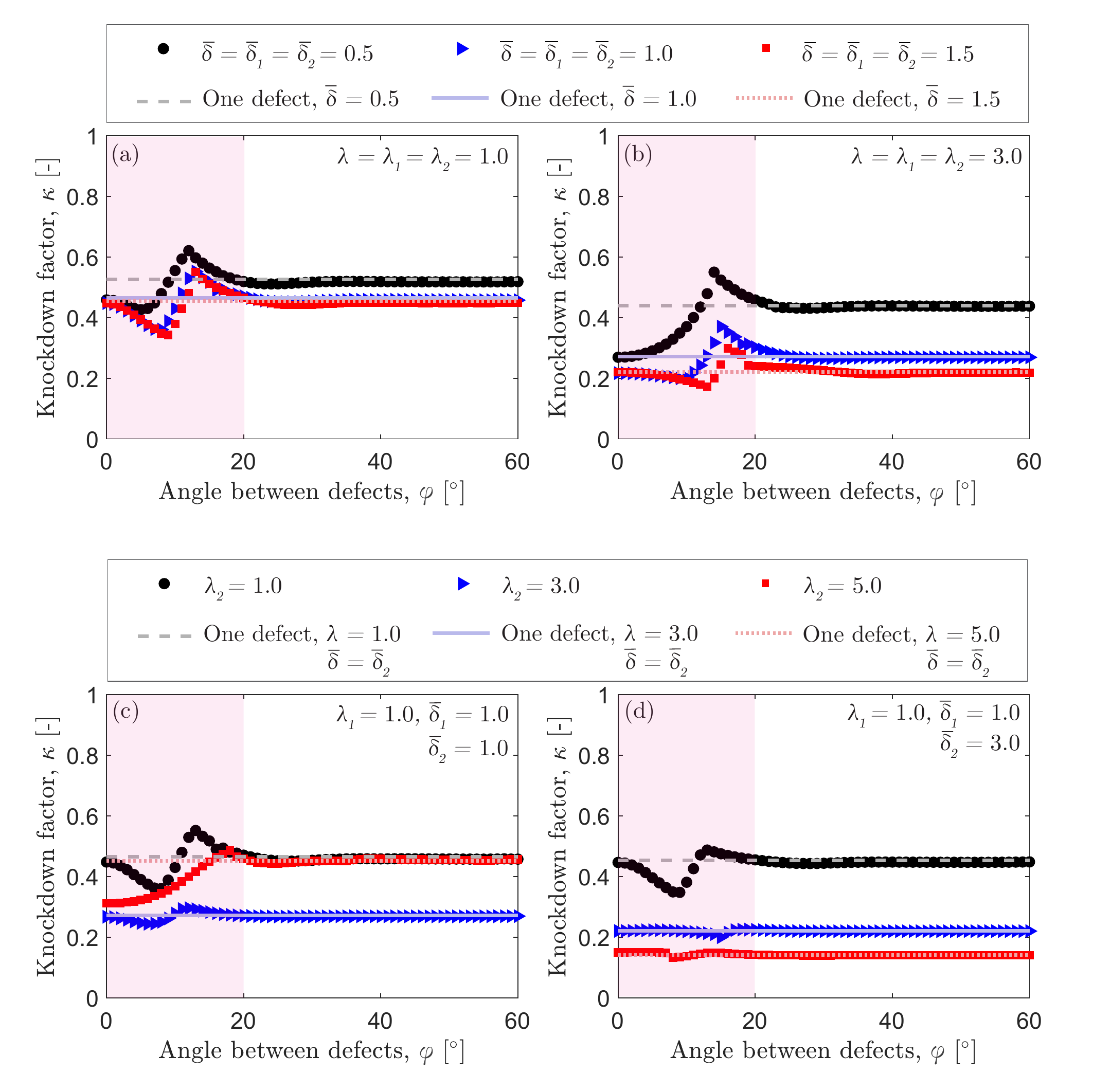}
    \caption{Knockdown factors, $\kappa$, versus angular separation between defects, $\varphi$, for imperfect shells containing two imperfections ($N=2$). The shaded areas represent the region where the defects interact.
    (a, b) Two \textit{identical} defects with constant normalized width (a) $\lambda=\lambda_{1}=\lambda_{2}=1$ or (b) $\lambda=\lambda_{1}=\lambda_{2}=3$,  and varying amplitude $\overline{\delta}$ (see legend). (c, d) Two \textit{distinct} defects with constant $(\overline{\delta}_{1},\, \lambda_{1})=(1,\,1)$, but (c) $\overline{\delta}_{2}=1$, or (d) $\overline{\delta}_{2}=3$, while varying $\lambda_{2}$ at each shell (see legend). The horizontal lines represent the knockdown factors for shells containing a single defect with $(\overline{\delta},\,\lambda)=(\overline{\delta}_2,\,\lambda_2)$.}
    \label{fig:twodefects}
    \end{figure}

First, we consider two identical defects (with the same amplitudes $\overline{\delta}=\overline{\delta}_1=\overline{\delta}_2$ and normalized widths $\lambda=\lambda_1=\lambda_2$). The angular 
distance, $\varphi$, between the two defects is varied systematically while setting one at the pole ($\beta_1=0$) and the other at $\beta_2=\varphi$. In the panels (a) and (b) of Fig.~\ref{fig:twodefects}, we plot the knockdown factor, $\kappa$, as a function of $\varphi$ for the cases of relatively narrow ($\lambda = 1$) and wide ($\lambda = 3$) defects, respectively. Three values of defect amplitude are considered: $\overline{\delta}=\{0.5,\,1.0,\,1.5 \}$ 
(see legend). As $\varphi$ increases, $\kappa$ first decreases to a minimum (the shell weakens) and then increases (the shell strengthens), surpassing the initial value of $\kappa(\varphi=0)$ to reach a maximum. After this maximum, $\kappa$ asymptotes to a constant for larger values of $\varphi$. This non-monotonic dependance of the buckling strength can be attributed to cross-interactions when the defects are nearby. Important interactions occurs for $\varphi \lesssim 20^\circ$ (shaded regions in Fig.~\ref{fig:twodefects}a,\,b), whereas a plateau with constant $\kappa$ is observed for $\varphi \gtrsim 20^\circ$. Within the interaction regime, there are both weakening effects (near $\varphi\approx 0$) and strengthening effects (for intermediate angular distances, $12^\circ \lesssim \varphi \lesssim 20^\circ$) with respect to the equivalent one-defect case. Recalling the shape of the individual Gaussian dimple  assumed in Eq.~(\ref{eqn:geom_gaussiandimple}), the combined profile when $N=2$ is
\begin{equation}\label{eqn:2defect_profile}
w(\beta,\,\theta) =  w_1(0,\,0)+w_2(\varphi,\,\theta_2).
\end{equation}
Near $\varphi\approx 0$, the weakening effect can be attributed to the near superposition of the profiles of the two defects, with an amplitude that is nearly twice that of the individual defects. The strengthening effect can be attributed to the potential restrain of the expected buckling mode of one of the defects by the other. 
Importantly, for $\varphi > 20^\circ$, the plateau of constant $\kappa$ coincides with that of a shell containing a single defect of the same $\overline{\delta}$ and $\lambda$ (solid, dashed, and dotted horizontal lines in Fig.~\ref{fig:twodefects}; see legend), as studied previously in Ref.~\cite{lee_geometric_2016}. Thus, when the two defects are sufficiently far apart, their interaction is negligible, and one of them acts as the \textit{weakest link} to govern the buckling capacity of the shell.

We now turn to the case when the two defects are different; the geometry of the $i=1$ defect is kept constant with $\lambda_1 = 1$ and $\overline{\delta}_1 = 1$, while the $i=2$ defect has  $\overline{\delta}_2=\{1,\,3\}$ and $\lambda_2=\{ 1,\,3,\,5\}$. In Fig.~\ref{fig:twodefects}(c,\,d), we present  $\kappa(\varphi)$ curves for shells with $\overline{\delta}_2=1$, $\overline{\delta}_2=3$ and $\lambda_2 = \{1,\,3,\,5\}$, respectively. 
Even if less prominently than for the data in Fig.~\ref{fig:twodefects}(a,\,b), we still find a non-monotonic region of interaction with strengthening and weakening effects, past which a constant plateau is reached for all curves. Coincidentally, the value of $\varphi \approx 20^\circ$ for the angular width separating the interaction region and the plateau, is similar to the identical-defects case. We do not believe that this is a general result. It is important to note that, for the two-defects case with different amplitudes ($\overline{\delta}_1=1,\overline{\delta}_2=3$) plotted in Fig.~\ref{fig:twodefects}(d), the $\kappa(\varphi)$ curves becomes nearly flat with increasing width of the $i=2$ defect ($\lambda_2 = \{3,\,5\}$), wider than $\lambda_1 = 1$ for the first defect, even for $\varphi \lesssim 20^\circ$. This result indicates the defects tend to interact more prominently when they have similar geometries.

In both of the $N=2$ cases explored above (identical and different defects), the exact values of the knockdown factor in the interaction region (strengthening and weakening) and in the plateau depend on the specific geometric properties of the two defects. Developing predictive knowledge for these nontrivial defect-defect interactions deserves a detailed investigation of its own, which is, however, beyond the scope of the present study. Still, the most important feature to retain from the above results is that the values of the plateau are dictated by the strongest defect, highlighting the dominance of the \textit{weakest link} in dictating the buckling of the shells when defects are sufficiently far apart but with more nontrivial interactions when they are nearby.

\section{Buckling of imperfect shells containing a distribution of defects}
\label{sec:results_distribution}

\begin{figure}[!h]
    \centering\includegraphics[width=0.84\textwidth]{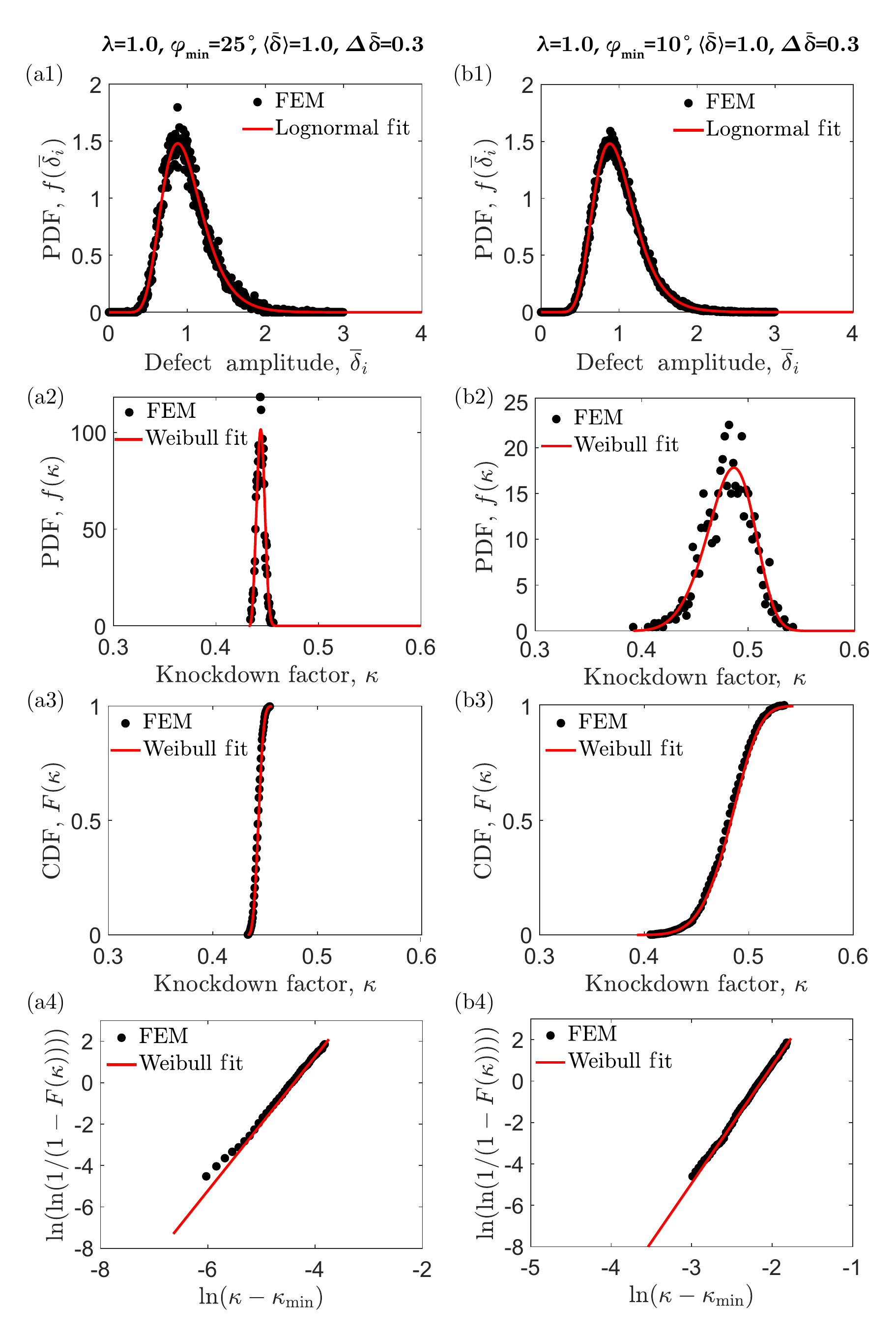}
    \caption{Probabilistic buckling of a shell containing a distribution of defects. (a1,\,b1) Input probability density functions (PDFs) of the amplitude of defects, $f(\overline{\delta}_{i})$, used for the shell design. (a2,\,b2) Output PDFs of knockdown factor, $f(\kappa)$. (a3,\,b3) Cumulative distribution functions (CDFs) of $\kappa$, $F(\kappa)$. (a4,\,b4) Weibull plots of $F(\kappa)$. The minimum angular separation between any two defects is set to $\varphi_\mathrm{min}=25^\circ$ in panels (a1-a4) and to $\varphi_\mathrm{min}=10^\circ$ in panels (b1-b4). All other design parameters are kept fixed at $\lambda=1$, $\overline{\delta} = 1.0$ and $\Delta \overline{\delta} = 0.3$. The PDFs in panels (a1,\,b1) were constructed using 500 bins. The PDFS in panels (a2-a4) were constructed with 2000 bins, and those in (b2-b4) with 500 bins, due to their different span across the considered range $0.3 \leq \kappa \leq 0.6$. 
    }
    \label{fig:histograms}
    \end{figure}

We are now ready to tackle our central problem, defined in Section~\ref{sec:problem_definition}, of the buckling of shells with a large number of imperfections ($N\gg1$). These shells contain defects, each of amplitude $\overline{\delta}_i$, which is distributed lognormally according to the PDF, $f(\overline{\delta}_i)$, in Eq.~(\ref{eqn:lognormal}), with mean amplitude $\langle \overline{\delta}\rangle$ and standard deviation $\Delta\overline{\delta}$. The explored sets of these parameters are:
$\langle \overline{\delta}\rangle =\{0.2,\,0.5,\,1,\,1.5,\,2,\,2.5,\,3\}$ and $\Delta \overline{\delta}=\{0,\,0.1,\,0.3,\,0.6,\,1\}$. For each shell design, we fix the defect width $\lambda=\{1,\,2\}$ and the minimum angular separation between any two defects $\varphi_\mathrm{min}=\{10^\circ,\,25^\circ\}$. Recalling that the threshold for defect-defect interactions in the $N=2$ case is $\varphi_\mathrm{min}\approx20^\circ$ (cf. Fig.~\ref{fig:twodefects}), the two values of $\varphi_\mathrm{min}$ were chosen to explore configurations where defect-defect interactions are expected to be negligible (for $\varphi_\mathrm{min}=25^\circ$) or important (for $\varphi_\mathrm{min}=10^\circ$). Seeding is done within a spherical cap with $\beta_\mathrm{max}=60^\circ$ to avoid boundary effects. 
For each set of design parameters $(\langle \overline{\delta}\rangle,\,\Delta \overline{\delta},\,\lambda,\,\varphi_\mathrm{min})$, we typically construct 200 realizations of statistically equivalent shell geometries, yielding a total of 28,000 FEM simulations. Only for the data in Fig.~\ref{fig:histograms}, we generated 1200 realizations per shell design to enhance the statistics and check for independence of the ensemble size.

In Fig.~\ref{fig:histograms}(a1), we show an example of the \textit{input} statistics for a design with $(\langle \overline{\delta}\rangle,\,\Delta \overline{\delta},\,\lambda,\,\varphi_\mathrm{min})=(1.0,\,0.3,\,1,\,25^\circ)$. We perform FEM simulations for each of the 1200 statistically equivalent designs, measure the corresponding knockdown factors, $\kappa$, and construct the \textit{output} PDF, $f(\kappa)$, shown in 
Fig.~\ref{fig:histograms}(a2). We find that the histogram obtained from the FEM data is described well by a 3-parameter Weibull distribution~\cite{weibull_statistical_2021} (solid line):
\begin{equation}\label{eqn:weibull}
f(\kappa)=\frac{\gamma}{\tilde{\kappa}}\left(\frac{\kappa - \kappa_\mathrm{min}}{\tilde{\kappa}} \right)^{\gamma-1}\,\exp\left(-\left(\frac{\kappa - \kappa_\mathrm{min}}{\tilde{\kappa}}\right)^\gamma\right) ,
\end{equation}
where $\tilde{\kappa}$, $\gamma$ and $\kappa_\mathrm{min}$ are the scale, shape, and location (threshold) parameters, respectively. 
The third fitting parameter, $\kappa_\mathrm{min}$, is required to account for the lower bound of $\kappa$, associated with the plateau of the $\kappa(\overline{\delta})$ curves, which was investigated in Ref.~\cite{jimenez2017technical} for single-defect shells and found to depend on $\lambda$ and $\eta$. In Fig.~\ref{fig:histograms}(b1), a second example with $(\langle \overline{\delta}\rangle,\,\Delta \overline{\delta},\,\lambda,\,\varphi_\mathrm{min})=(1.0,\,0.3,\,1,\,10^\circ)$ also yields Weibull statistics for the knockdown factor (Fig.~\ref{fig:histograms}b2). 

The three Weibull fitting parameters ($\tilde{\kappa}$, $\gamma$, $\kappa_\mathrm{min}$) used to plot the Weibull distribution in Fig.~\ref{fig:histograms}(a2) and~\ref{fig:histograms}(b2) were obtained based on the Maximum Likelihood Estimates (\texttt{*mle} function in Matlab) and determined to be ($0.012$, $3.242$, $0.433$) and ($0.119$, $5.663$, $0.372$), respectively. Using these fitting parameters, we also computed the corresponding Weibull cumulative distribution functions (CDFs),
\begin{equation}\label{eqn:weibull_cdf}
F(\kappa)=1-\,\exp\left(-\left(\frac{\kappa - \kappa_\mathrm{min}}{\tilde{\kappa}}\right)^\gamma\right),
\end{equation}
which are in excellent agreement with the histograms obtained from the FEM data, as shown in Fig.~\ref{fig:histograms}(a3) and Fig.~\ref{fig:histograms}(b3). As a double-check, if the FEM data for the $F(\kappa)$ statistics is indeed represented by a Weibull CDF, plotting $\ln{(\ln{(1/(1-F(\kappa)))})}$ as a function of $\ln{(\kappa-\kappa_{\mathrm{min}})}$ is expected to yield a straight line of slope $\gamma$, which is confirmed in Fig.~\ref{fig:histograms}(a4) and Fig.~\ref{fig:histograms}(b4) for the $\varphi_\mathrm{min}=25^\circ$ and $\varphi_\mathrm{min}=10^\circ$ cases, respectively. 

The Weibull distribution functions in Eq.~(\ref{eqn:weibull}) and Eq.~(\ref{eqn:weibull_cdf}) are derived based on extreme value theory~\cite{fisher1928limiting}, under the assumption that the failure probability of one representative element of a structure follows a power-law tail~\cite{le2015modeling}. Within this framework, the failure of one of the elements, the \textit{weakest link}, yields the global failure of the structure~\cite{weibull1939phenomenon,weibull1951statistical,bavzant2009scaling,le2009strength,le2015modeling}. The above observations from the data in Fig.~\ref{fig:histograms} indicate the suitability of the 3-parameter Weibull distribution to describe the statistics of the knockdown factors of shells with lognormally distributed defect amplitudes, suggesting that probabilistic shell buckling can be placed within the class of extreme-value phenomena.

    \begin{figure}[!h]
    \centering\includegraphics[width=0.81\textwidth]{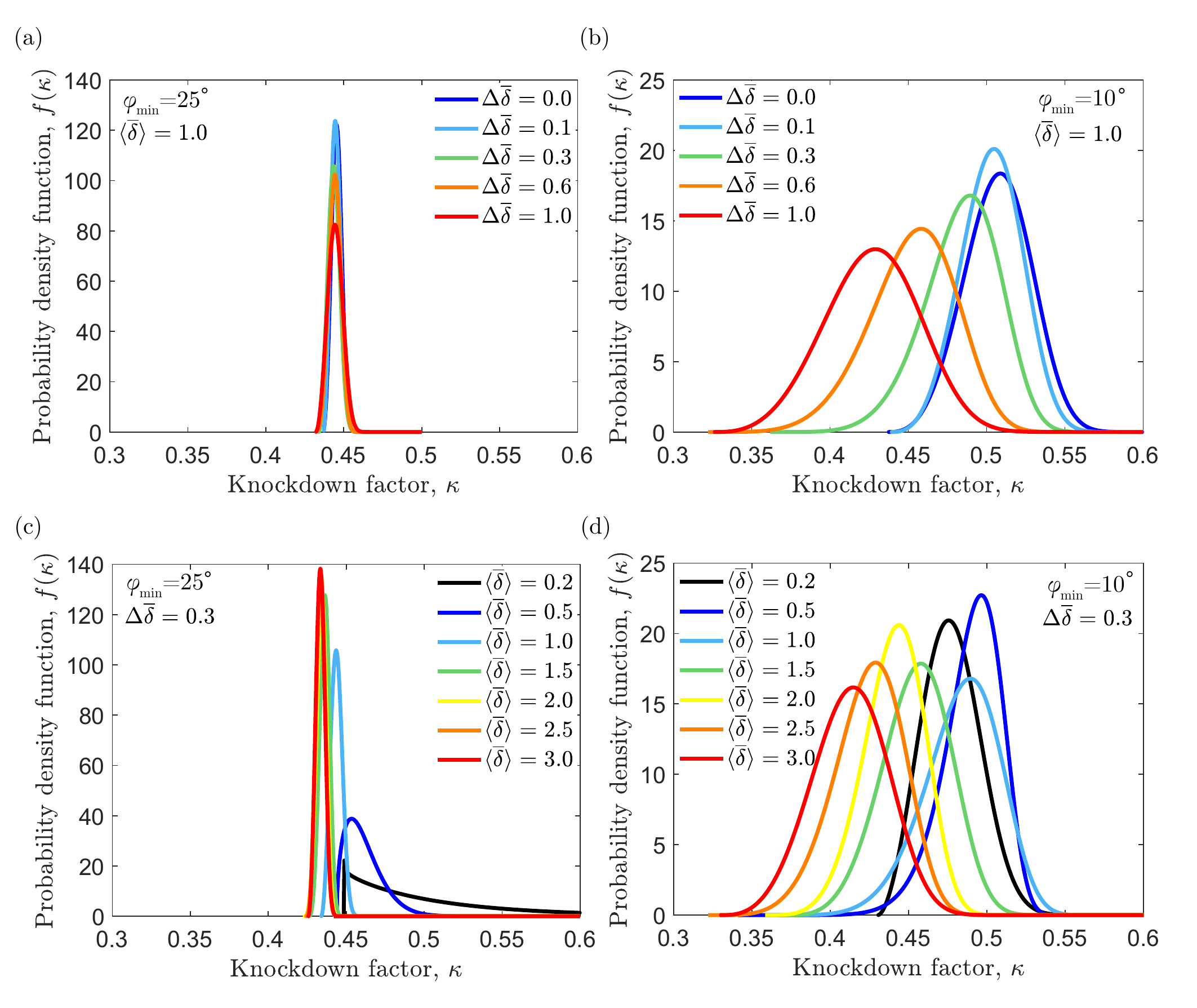}
    \caption{Probability density functions of the knockdown factor, $f(\kappa)$, obtained by fitting Eq.~(\ref{eqn:weibull}) to the FEM data, for a variety of design parameters. (a,\,b) Constant amplitude $\langle\overline{\delta}\rangle = 1$ and varying $0 \leq \Delta \overline{\delta} \leq 1.0$ (see legend); for (a) $\varphi_\mathrm{min}=25^\circ$ and (b) $\varphi_\mathrm{min}=10^\circ$. (c,\,d) Constant value of $\Delta \overline{\delta} =0.3$ and varying $ 0.2 \leq \langle\overline{\delta}\rangle \leq 3.0$; for (c) $\varphi_\mathrm{min}=25^\circ$ and for (d) $\varphi_\mathrm{min}=10^\circ$.}
    \label{fig:weibull_pdfs}
    \end{figure}

In Fig.~\ref{fig:weibull_pdfs}, we now plot a set of PDFs, $f(\kappa)$, obtained by fitting the FEM data similarly to what was done in Fig.~\ref{fig:histograms}, for a wider range of the parameters $\langle \overline{\delta} \rangle$ and $\Delta \overline{\delta}$. For clarity, we only show the fitted PDFs and not the actual histograms of the FEM data. The data in Fig.~\ref{fig:weibull_pdfs}(a,\,b) explore various $\Delta \overline{\delta}$, $0.0 \leq \Delta \overline{\delta} \leq 1.0$ (while fixing $\langle \overline{\delta} \rangle = 1.0$). In Fig.~\ref{fig:weibull_pdfs}(c,\,d), we explore various $\langle \overline{\delta} \rangle$, $0.2 \leq \langle \overline{\delta} \rangle \leq 3.0$ (while fixing $\Delta \overline{\delta}=0.3$). Panels (a,\,c) are for $\varphi_\mathrm{min}=25^\circ$ and panels (b,\,d)  are for $\varphi_\mathrm{min}=10^\circ$. 
First, the data in Fig.~\ref{fig:weibull_pdfs}(a), with fixed $\langle \overline{\delta} \rangle = 1.0$, shows that the mode (location of the peak) of $f(\kappa)$ remains approximately constant, for all $\Delta \overline{\delta}$, even if the peak probability decreases slightly with $\Delta \overline{\delta}$. In Fig.~\ref{fig:weibull_pdfs}(b), when the minimum defect-to-defect distance is decreased to $\varphi_\mathrm{min}=10^\circ$, we find that $\kappa$ decreases with $\Delta \overline{\delta}$, presumably due to the higher probability of randomly seeding defect amplitudes from the high tail of $f(\overline{\delta}_{i})$, coupled with defect-defect interactions. Furthermore, the fitted Weibull threshold decreases consistently with increasing $\Delta \overline{\delta}$. Similarly, when fixing $\Delta \overline{\delta}=0.3$ and increasing the defect amplitude $\langle \overline{\delta} \rangle$ (Fig.~\ref{fig:weibull_pdfs}c,\,d), we find a modest decrease in $\kappa$ for $\varphi_\mathrm{min}=25^\circ$ (Fig.~\ref{fig:weibull_pdfs}c), where the strongest defect governs most of the shell behavior. For $\varphi_\mathrm{min}=10^\circ$ (Fig.~\ref{fig:weibull_pdfs}d), this behavior is more pronounced due to defect-defect interactions.

\begin{figure}[!h]
    \centering\includegraphics[width=0.81\textwidth]{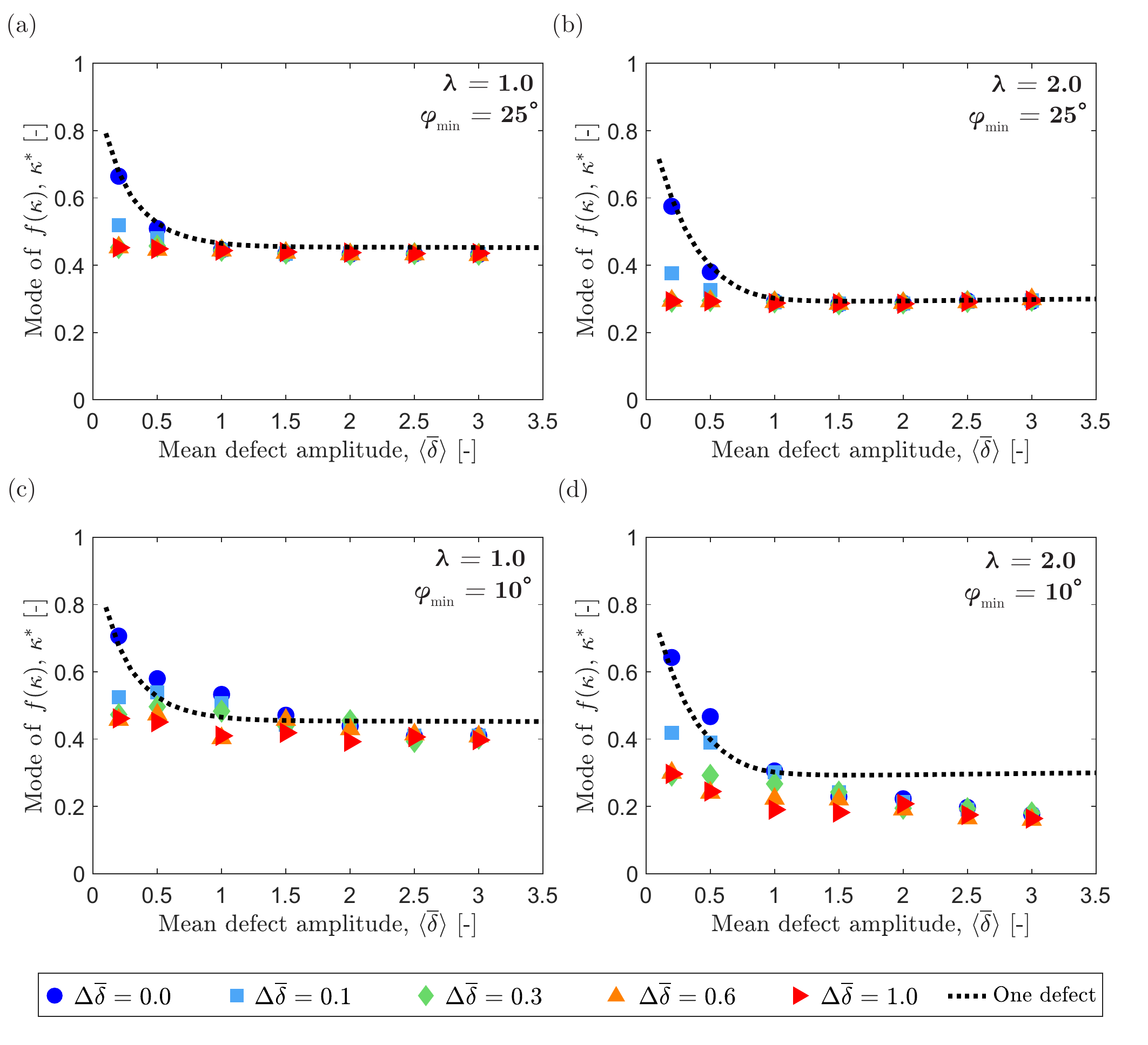}
    \caption{Mode of knockdown factors, $\kappa^{*}$, versus the mean defect amplitude, $ \langle \overline{\delta} \rangle$. (a,\,b) $\varphi_\mathrm{min} = 25^\circ$ (non-interacting defects) and widths $\lambda=1$ and $\lambda=2$, respectively. (c,\,d) $\varphi = 10^\circ$ (interacting defects) and widths $\lambda=1$ and $\lambda=2$, respectively. The various datasets correspond to different values of $\Delta \overline{\delta}$ (see legend), and are compared to the equivalent single-defect cases (dotted lines).} 
    \label{fig:knockdowns1}
    \end{figure}

It is of interest to quantify the mode, $\kappa^{*}$; \textit{i.e.} location of the peak of $f(\kappa)$, for all of our data. In Fig.~\ref{fig:knockdowns1}, we plot $\kappa^{*}$ as a function of the mean defect amplitude, $\langle \overline{\delta}\rangle$, for four different design configurations. Panels (a,\,b) of Fig.~\ref{fig:knockdowns1} correspond to $\varphi_\mathrm{min}=25^\circ$, and panels (c,\,d) to $\varphi_\mathrm{min}=10^\circ$. Also, panels (a,\,c) are for $\lambda=1.0$, and panels (b,\,d) for wider defects with $\lambda=2.0$. The various datasets (see legend) have different values of $\Delta \overline{\delta}$. In all plots, the dotted lines represent results for a shell with a single defect of amplitude $\langle \overline{\delta}\rangle$. This FEM data for one-defect shells were verified with the data in Ref.~\cite{lee_geometric_2016}.

Focusing first on Fig.~\ref{fig:knockdowns1}(a), we find that the mode of the knockdown factor, $\kappa^{*}$, decreases with increasing $\Delta \overline{\delta}$, due to the higher probability of seeding deeper, and hence more dominant, defects in the high tail of $f(\overline{\delta}_i)$ for higher values of $\Delta \overline{\delta}$. This decrease of $\kappa^{*}$ is more pronounced when $\langle\overline{\delta}\rangle\lesssim 1$ and less so in the plateau region, for $\langle\overline{\delta}\rangle >1$. The plateau has an approximate constant value $\kappa^{*}\approx0.45$, which is in agreement with predictions of the analogous plateau for single-defect shells~\cite{jimenez2017technical}. 
These findings suggest that the buckling capacity of  shells containing a distribution of imperfections is dominated by the deepest defect, their \textit{weakest link}. This scenario is qualitatively similar when $\lambda=2$, as shown in Fig.~\ref{fig:knockdowns1}(b), where the plateau region also matches the single-defect case. These various features are slightly different for the datasets in Fig.~\ref{fig:knockdowns1}(c, d), when $\varphi = 10^\circ$. Here, for example, the dataset with $\Delta \overline{\delta}=0$ (circles) does not coincide with the one-defect curve (dotted line), especially for $\lambda=2$ (Fig.~\ref{fig:knockdowns1}d), which can be attributed to defect-defect interactions. Moreover, when $\varphi = 10^\circ$, the resultant plateau of $\kappa^*$ is consistently below that of shells with a single defect~\cite{jimenez2017technical}, especially in Fig.~\ref{fig:knockdowns1}(d), further indicating the importance of defect-defect interactions in this regime.

\begin{figure}[!h]
    \centering\includegraphics[width=0.81\textwidth]{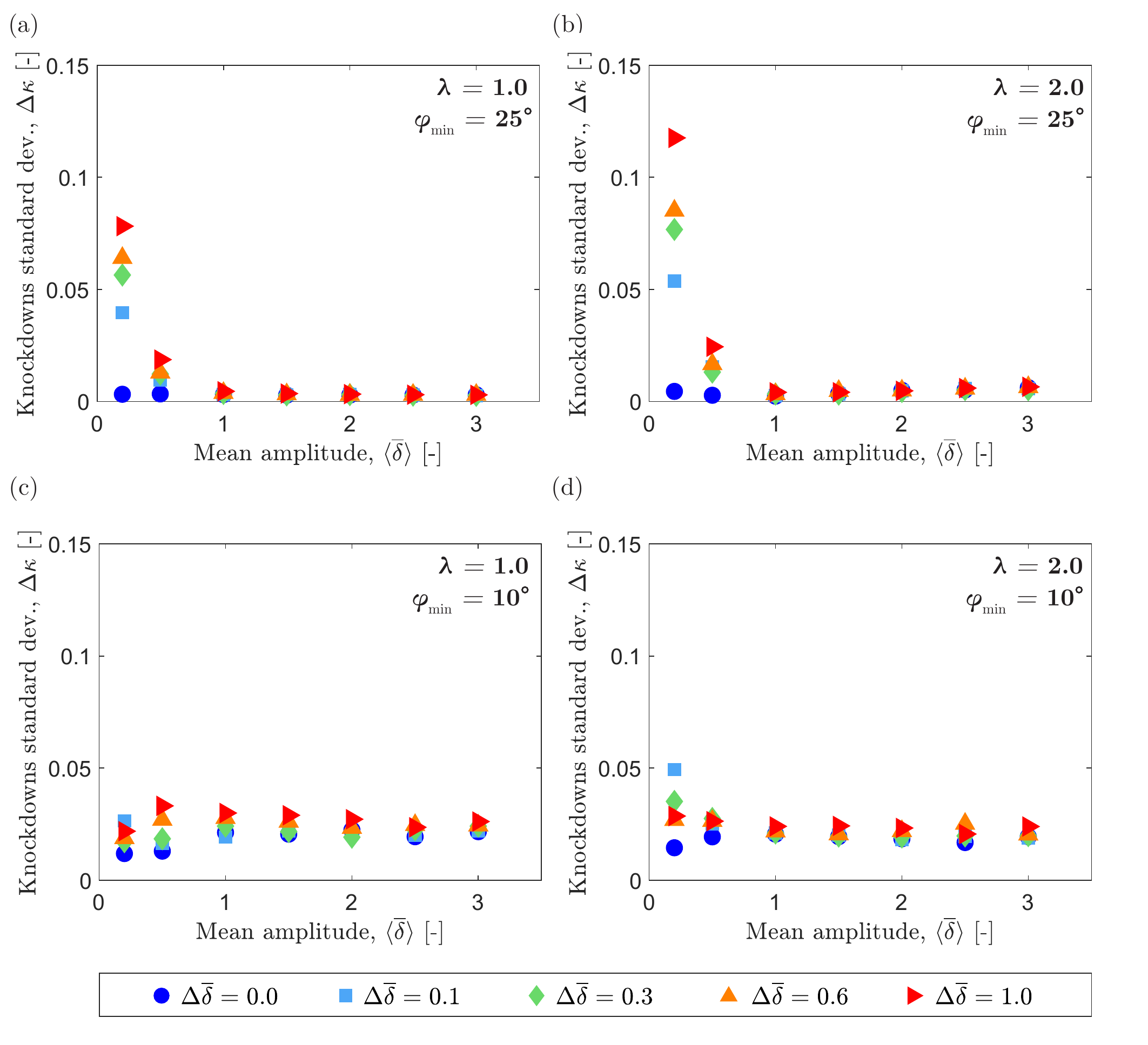}
    \caption{Standard deviation of the resultant knockdown factors, $\Delta \kappa$, versus the mean defect amplitude, $ \langle \overline{\delta} \rangle$. (a,\,b) $\varphi_\mathrm{min} = 25^\circ$ (non-interacting defects) and widths $\lambda=1$ and $\lambda=2$, respectively. (c,\,d) $\varphi = 10^\circ$ (interacting defects) and widths $\lambda=1$ and $\lambda=2$, respectively. The various datasets correspond to different values of $\Delta \overline{\delta}$ (see legend).} 
    \label{fig:knockdownsstds}
    \end{figure}
    
Returning to the PDFs presented in Fig.~\ref{fig:weibull_pdfs}, we observe qualitatively that the $f(\kappa)$ distributions are typically narrow for $\varphi = 25^\circ$ (when there are \textit{negligible} defect-defect interactions; \textit{e.g.}, Fig.~\ref{fig:weibull_pdfs}a) and broad for $\varphi = 10^\circ$ (when there are \textit{significant} defect-defect interactions; \textit{e.g.}, Fig.~\ref{fig:weibull_pdfs}b). 
In the final step of our investigation, we seek to quantify the width of the $f(\kappa)$ distributions, as measured by the standard deviation of the resultant knockdown factor, $\Delta \kappa$, for a variety of design parameters. The corresponding data are shown in Fig.~\ref{fig:knockdownsstds}.
With \textit{negligible} defect-defect interactions ($\varphi = 25^\circ$; see Fig.~\ref{fig:knockdownsstds}a,\,b), $f(\kappa)$ remains consistently narrow ($\Delta \kappa$ is small) when $\langle \overline{\delta} \rangle \gtrsim 1$, but broadening occurs for $\langle \overline{\delta} \rangle \lesssim 1$, significantly more so for the larger values of $\Delta \overline{\delta}$. This behavior is robust for the two values of $\lambda$ explored in Fig.~\ref{fig:knockdownsstds}(a, b). By contrast, when there are \textit{important} defect-defect interactions ($\varphi = 10^\circ$; see Fig.~\ref{fig:knockdownsstds}c, d), $f(\kappa)$ is always broad across most of the range of $\langle \overline{\delta} \rangle$, and fairly independent of $\Delta \overline{\delta}$. 
From the perspective of structural reliability,
these results highlight the following possible interpretation scenario. Shells with non-interacting defects with large amplitudes ($\langle \overline{\delta} \rangle \gtrsim 1$) appear to exhibit quasi-deterministic buckling, with narrow $f(\kappa)$ distributions. By contrast, shells with either small-amplitude defects ($\langle \overline{\delta} \rangle \lesssim 1$) or interacting defects exhibit a far more probabilistic behavior, with broad $f(\kappa)$ distributions, and have, therefore, significantly lower reliability.


\section{Discussions and Conclusions}
\label{sec:conclusion}

We have employed experimentally validated FEM simulations to investigate the buckling of spherical shells containing a random distribution of defects, seeking to quantify the resultant knockdown factor ($\kappa$) statistics, as measured by the probability density function, $f(\kappa)$. First, we used 3D-printed molds and a polymer coating technique to fabricate imperfect hemispherical shells containing distributions of defects. The imperfections comprised \textit{outward} defects (bumps) and the full geometry of the shell was characterized through $\mu$CT. Using the $\mu$CT geometric data, a high-fidelity 3D shell FEM was validated against experimental buckling measurements. 

After validating the FEM, we first focused on imperfect shells containing only two defects ($N=2$) to characterize the influence of defect-defect interactions on the buckling conditions. The results showed these interactions can be significant when the angular separation between the two defects is below a threshold value ($\varphi \lesssim 20^\circ$). Within this regime of interactions, we observed both weakening (near $\varphi\approx 0$) and strengthening (for intermediate angular distances, $12^\circ \lesssim \varphi \lesssim 20^\circ$) effects. When the defects are further apart, outside of the interactions regime ($\varphi\gtrsim 20^\circ$), the knockdown factor tends to a constant value, coinciding with the equivalent one-defect case. In the absence of defect-defect interactions, this result indicates that the shell buckling is dictated by the strongest defect of the pair -- its \textit{weakest link}.

We then focused on the central part of our study, where multiple defects ($N\gg1$) were distributed randomly on the surface of the spherical shell. The amplitude $\overline{\delta}_{i}$ of each defect was treated as a random variable sampled from a lognormal distribution of mean $\langle \overline{\delta}\rangle $, and standard deviation $\Delta \overline{\delta}$, while fixing the defect width and the minimum defect-defect angular separation for each shell design. In what we see as the most important contribution of this work, we find that, given a set of design parameters of shells containing many defects (whose amplitudes are lognormally distributed), the statistics of the resultant knockdown factor are described by the 3-parameter Weibull distribution, $f(\kappa)$ in Eq.~(\ref{eqn:weibull}). This result is consistent with other extreme-value statistics problems where, in their general form, an \textit{input} distribution of \textit{links}, whose individual failure probability follows a power-law tail, yields an \textit{output} Weibull distribution \cite{jayatilaka_statistical_1977,le2015modeling}. Further analyzing the resultant Weibull distribution, for all input $\langle \overline{\delta}\rangle$ and $\Delta \overline{\delta}$, we found that the output distributions are consistently narrow when there are less prominent interactions between defects ($\varphi_\mathrm{min}=25^\circ$), in comparison to the broader distributions when defect-defect interactions are present ($\varphi_\mathrm{min}=10^\circ$). These findings for the width of $f(\kappa)$, together with the results for its mode, are consistent with the \textit{weakest-link} interpretation where the deepest defect governs the global shell buckling.

In future work, using the approach reported here, we intend to revisit and rationalize the results from the seminal experiments by Carlson \textit{et al.}~\cite{carlson1967experimental}, where increasing knockdown factors were observed when progressively eliminating severe defects. More broadly, we believe that our findings will open an exciting avenue for future study on probabilistic shell buckling, including theoretical methods based on extreme-value statistics and weakest-link models.
\bigskip

\noindent \textbf{Acknowledgments:} The authors are grateful to Roberto Ballarini and Jia-Liang Le for discussions on extreme-value statistics and their potential connection to shell buckling, which motivated the present study. We also thank Arefeh Abbasi and Luna Lin for a critical read of the present manuscript and for their valuable feedback. 

\bibliography{Bibliography.bib}

\end{document}